\documentclass[preprint,aps,prd,superscriptaddress,showpacs,nofootinbib]{revtex4}%
\usepackage{graphicx}
\usepackage{amsmath}
\usepackage{amssymb}
\usepackage{bm}
\usepackage{epsfig}
\usepackage[normalem]{ulem}
\begin{document}
\preprint{TUM-EFT 123/19}
\preprint{P3H-19-016}
\preprint{TTP19-019}
\title{\mbox{}\\[10pt]
Order $\bm{v^4}$ corrections to Higgs boson decay into $\bm{J/\psi + \gamma}$
}
\author{Nora~Brambilla}
\affiliation{Physik-Department, Technische Universit\"at M\"unchen,
James-Franck-Str. 1, 85748 Garching, Germany}
\affiliation{Institute for Advanced Study, Technische Universit\"at M\"unchen,
Lichtenbergstrasse 2~a, 85748 Garching, Germany}
\author{Hee~Sok~Chung}
\affiliation{Physik-Department, Technische Universit\"at M\"unchen,
James-Franck-Str. 1, 85748 Garching, Germany}
\author{Wai~Kin~Lai}
\affiliation{Physik-Department, Technische Universit\"at M\"unchen,
James-Franck-Str. 1, 85748 Garching, Germany}
\author{Vladyslav Shtabovenko}
\affiliation{Zhejiang  Institute of Modern Physics, Department of Physics, 
Zhejiang University, Hangzhou 310027, China}
\affiliation{Institut f\"ur Theoretische Teilchenphysik (TTP), 
Karlsruher Institut f\"ur Technologie (KIT), 76131 Karlsruhe, Germany}
\author{Antonio~Vairo}
\affiliation{Physik-Department, Technische Universit\"at M\"unchen,
James-Franck-Str. 1, 85748 Garching, Germany}
\date{\today}
\begin{abstract}
The process $H \to J/\psi + \gamma$, where $H$ is the Higgs particle,
provides a way to probe the size and the sign of the Higgs-charm coupling. 
In order to improve the theoretical control of the decay rate, 
we compute order $v^4$ corrections to the decay rate  
based on the nonrelativistic QCD factorization formalism. 
The perturbative calculation is carried out by using automated computer codes. 
We also resum logarithms of the ratio of the masses of the Higgs boson and the 
$J/\psi$ 
to all orders in the strong coupling constant $\alpha_s$ to next-to-leading
logarithmic accuracy. 
In our numerical result for the decay rate, 
we improve the theoretical uncertainty, 
while our central value is in agreement with previous studies within errors. 
We also present numerical results for $H \to \Upsilon(nS) + \gamma$ for
$n=1,2$, and 3, 
which turn out to be extremely sensitive to the Higgs bottom coupling.
\end{abstract}
\maketitle
\section{Introduction}
\label{sec:introduction}%

The investigation of the Higgs sector of the Standard Model is one of the most 
important areas of particle physics today. 
While measuring the Higgs boson self-couplings
will reveal important information about electroweak symmetry breaking, 
the determination of the Yukawa couplings between the Higgs $H$ and the
Standard Model fermions 
is a direct probe of the origin of fermion masses. 
While current measurements of Higgs production at the LHC provide some 
constraint on the Higgs top and Higgs bottom Yukawa
couplings~\cite{Khachatryan:2016vau}, 
a determination of the Higgs charm coupling is still out of reach. 

The possibility of measuring the Higgs charm coupling at the high-luminosity
LHC (HL-LHC) has been studied in two different processes~\cite{Aaboud:2018txb, 
Aaboud:2018fhh}.
One way is to measure the 
Higgs decay into $c \bar c$, by identifying charm jets in the final state. 
Another way is to measure the decay of the Higgs boson to a charmonium and a
photon~\cite{Bodwin:2013gca}. 
Compared to $H \to c \bar c$, the process $H \to$ charmonium$+ \gamma$ has
an advantage that the 
charmonium provides a clean final state through its electromagnetic decays.  
Higgs decay into charmonium$+ \gamma$ also allows a simultaneous measurement of
the size and the sign of the Higgs charm coupling. 
The current upper limits for the branching ratio
${\rm Br} (H \to J/\psi + \gamma)$ and the cross section 
$\sigma (pp \to ZH) \times {\rm Br} (H \to c \bar c)$ at 95\% confidence level
are both about 2 orders of magnitude larger than the Standard-Model
predictions~\cite{Aaboud:2018txb, Aaboud:2018fhh}.

It is crucial that the decay rate $\Gamma (H \to J/\psi + \gamma)$
is in good theoretical control in order that the measurement of the rate leads 
to a determination of the Higgs charm coupling. Recently there have been 
many efforts to
improve the theoretical prediction of the decay rate within the Standard
Model~\cite{Bodwin:2013gca, Bodwin:2014bpa, Koenig:2015pha, Bodwin:2016edd,
Bodwin:2017wdu}.
Especially, approaches based on nonrelativistic effective field theories allow 
a systematic improvement of theoretical accuracy~\cite{Bodwin:2013gca, 
Bodwin:2014bpa, Bodwin:2016edd, Bodwin:2017wdu}.
In the nonrelativistic QCD (NRQCD) effective field theory~\cite{Bodwin:1994jh}, 
decay and production processes involving a heavy quarkonium are given by a 
double series in $\alpha_s$ and $v$, where $v$ is the typical velocity of a heavy
quark $Q$ in a heavy quarkonium; 
for charmonium, $v^2 \approx 0.3$, and for bottomonium, $v^2 \approx 0.1$.
Currently, the decay rates  
$\Gamma (H \to V + \gamma)$ for $V = J/\psi$ or $\Upsilon(nS)$ 
for $n=1,2,$ and $3$ 
have been computed to relative order $\alpha_s v^0$ and $v^2$ 
accuracy~\cite{Shifman:1980dk, Bodwin:2014bpa, 
Bodwin:2016edd, Bodwin:2017wdu}.
In Refs.~\cite{Bodwin:2014bpa, Bodwin:2016edd, Bodwin:2017wdu}, 
the large logarithms of $m_H^2/m_V^2$ that appear in higher order
corrections in $\alpha_s$, where $m_H$ is the Higgs mass
and $m_V$ is the mass of the quarkonium $V$, have been resummed to all orders
in $\alpha_s$ by combining the NRQCD and the light-cone
formalisms~\cite{Lepage:1980fj,Chernyak:1983ej, Jia:2008ep}.  

\begin{figure}
\epsfig{file=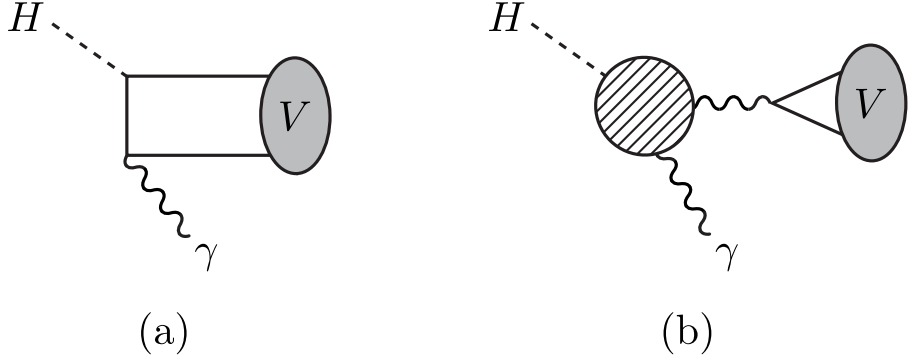,width=12cm}
\caption{
\label{fig:diags}%
Feynman diagrams for the (a) direct amplitude at order $\alpha_s^0$ and 
(b) the indirect amplitude for the process $H \to V + \gamma$. 
}
\end{figure}

In this paper, we improve the accuracy of the Standard Model 
prediction of the decay rates 
$\Gamma (H \to V + \gamma)$ for $V = J/\psi$ or $\Upsilon(nS)$ 
for $n=1,2,$ and $3$ 
by computing the order-$v^4$ correction to the decay rate in the NRQCD
factorization formalism. 
In our numerical analysis, we do not consider the $\psi(2S)$ meson,
because, to date, there are no available estimates of the relevant
NRQCD matrix elements accounting for open-flavor threshold effects and
nonrelativistic corrections in a complete and model-independent way. These
effects may be particularly important for this state, as it is just
$43$~MeV below the $D\bar{D}$ threshold.
We work in the limit $m_V^2 / m_H^2 \to 0$, where the calculation simplifies
dramatically. In this limit, $H \to V + \gamma$ occurs through two 
distinct processes that we refer to as direct and indirect processes; see
Fig.~\ref{fig:diags}.
In the direct process, the Higgs boson decays into a heavy quark $Q$
and a heavy antiquark $\bar Q$ through the Yukawa interaction, 
and the $Q \bar Q$ pair forms a
quarkonium after emitting a photon. We compute this amplitude to order-$v^4$
accuracy. 
We also resum the logarithms of $m_H^2/m_V^2$ to all orders in $\alpha_s$,
using the light-cone formalism, to next-to-leading logarithmic (NLL) 
accuracy.
That is, we resum the leading and next-to-leading logarithmic corrections
of the forms 
$\alpha_s^n \log^n (m_H^2/m_V^2)$ and 
$\alpha_s^n \log^{n-1} (m_H^2/m_V^2)$, respectively, 
for all orders $n \geq 1$ in $\alpha_s$. 
We note that the light-cone formalism applies only to the 
leading-order piece in the expansion in powers of 
$m_V^2/m_H^2$~\cite{Lepage:1980fj,Chernyak:1983ej}. 
In the indirect process, the Higgs boson
first decays into a $\gamma$ and a $\gamma^*$, and the $\gamma^*$ evolves into a
quarkonium~\cite{Bodwin:2013gca}. 
We compute the indirect amplitude to the same accuracy as the direct
amplitude. We compute the direct and indirect amplitudes separately because, in
the limit $m_V^2 / m_H^2 \to 0$, we find simplifications in the indirect
process that let us compute the indirect amplitude accurately from 
the known calculation of the $H \to \gamma \gamma$ decay amplitude and the 
leptonic decay rate of the meson $V$. 
Also, the logarithms of $m_H^2/m_V^2$ do not appear in the indirect
amplitude. 

We note that, although the indirect amplitude involves one photon
coupling more than the direct amplitude, this is compensated by the
heavy-quark-Higgs Yukawa coupling in the direct amplitude. 
In the case of the charm quark the Yukawa coupling is $y_c \approx 0.005$,
which indeed makes the direct amplitude numerically smaller than the indirect
one. In the case of the bottom quark $y_b \approx 0.018$ and the two amplitudes
are numerically close. See Sec.~\ref{sec:results}.

The remainder of this paper is organized as follows. 
In Sec.~\ref{sec:direct}, 
we compute the direct amplitude to relative order $v^4$ accuracy in the NRQCD
factorization formalism. 
We include the previously known order $\alpha_s$ and order $v^2$ corrections
and resum leading and next-to-leading 
logarithms of $m_H^2/m_V^2$ to all orders in $\alpha_s$. 
We compute the indirect amplitude in Sec.~\ref{sec:indirect}. 
We provide our numerical results in Sec.~\ref{sec:results}, and conclude in 
Sec.~\ref{sec:summary}.

\section{Calculation of the direct amplitude}
\label{sec:direct}%

In this section, we compute the direct amplitude to order $v^4$ accuracy in the
NRQCD factorization formalism. We work at leading order in $\alpha_s$, but we
will include the previously known order $\alpha_s v^0$ correction in our final
results. 

We first explain the formalism that we use to compute the direct amplitude in 
this section. 
The creation amplitude of a heavy quarkonium $V$ with polarization vector
$\epsilon(\lambda)$ and a photon
to relative order $v^4$ accuracy is given by 
\begin{eqnarray}
\label{eq:Jpsifactorization}%
i {\cal M}(H \to V + \gamma) &=& 
c_0 \langle V | \psi^\dag \bm{\sigma} \cdot \bm{\epsilon} (\lambda) \chi 
| 0 \rangle 
+ \frac{c_{\bm{D}^2}}{m^2} 
\langle V | \psi^\dag \bm{\sigma} \cdot \bm{\epsilon} (\lambda)
(-\tfrac{i}{2} \overleftrightarrow{\bm{D}})^2 \chi | 0 \rangle 
\nonumber\\ && 
+ \frac{c_{\bm{D}^4}}{m^4} \langle V | \psi^\dag \bm{\sigma} \cdot \bm{\epsilon} (\lambda)
(-\tfrac{i}{2} \overleftrightarrow{\bm{D}})^4 \chi | 0 \rangle 
\nonumber\\ && 
+ \frac{c_{\bm{D}^{(i} \bm{D}^{j)}}}{m^2} 
\langle V | \psi^\dag 
\epsilon^i (\lambda) \sigma^j (-\tfrac{i}{2})^2 
\overleftrightarrow{\bm{D}}^{(i} \overleftrightarrow{\bm{D}}^{j)} 
\chi | 0 \rangle 
\nonumber\\ && 
+ \frac{c_B}{m^2} \langle V | \psi^\dag g_s \bm{B} \cdot 
\bm{\epsilon} (\lambda)  \chi | 0 \rangle 
\nonumber\\ && 
+ \frac{c_{DE_0}}{m^3} 
\langle V | \psi^\dag 
\bm{\sigma} \cdot \bm{\epsilon} (\lambda) 
\tfrac{1}{3} 
( \overleftrightarrow{\bm{D}} \cdot g_s \bm{E} 
+g_s \bm{E} \cdot \overleftrightarrow{\bm{D}} ) 
\chi | 0 \rangle 
\nonumber\\ && 
+ \frac{c_{DE_1}}{m^3} 
\langle V | \psi^\dag \bm{\epsilon} (\lambda)  \cdot 
\tfrac{1}{2} 
[\bm{\sigma} \times ( \overleftrightarrow{\bm{D}} \times g_s \bm{E}
- g_s \bm{E} \times \overleftrightarrow{\bm{D}} )] \chi | 0 \rangle . 
\end{eqnarray}
Here, $m$ is the mass of the heavy quark, $g_s$ is the strong coupling, 
$\psi^\dag$ and $\chi$ are Pauli spinor fields that create a heavy quark
and an antiquark, respectively, 
and 
$E^i = G^{i0}$ and $B^i = \frac{1}{2} \epsilon^{ijk} G^{kj}$ are 
chromoelectric and chromomagnetic fields, respectively, where $G^{\mu \nu}$ is
the gluon field-strength tensor. 
The covariant derivative 
$\bm{D} = \bm{\nabla} - i g_s \bm{A}$ 
appears in Eq.~(\ref{eq:Jpsifactorization}) in the combination 
$\psi^\dag \overleftrightarrow{\bm{D}} \chi
 = \psi^\dag \bm{D} \chi - (\bm{D} \psi)^\dag \chi$. 
Operators with more than one covariant derivative are defined with 
\begin{eqnarray}
\label{eq:many_covariant_derivatives}%
\psi^\dag \overleftrightarrow{D}^{i_1} \ldots \overleftrightarrow{D}^{i_n} \chi
&=& (-1)^n \left( D^{i_1} \ldots
D^{i_n} \psi \right)^\dag \chi
\nonumber \\ && 
+ (-1)^{n-1}  
\left( D^{i_2} \ldots
D^{i_n} \psi \right)^\dag 
D^{i_1}\chi
\ldots 
+ \psi^\dag D^{i_1} \ldots D^{i_n} \chi.
\end{eqnarray}
The notation $T^{(ij)} = \tfrac{1}{2} (T^{ij}+T^{ji}) - 
\tfrac{1}{3} T^{ii} \delta^{ij}$
is a shorthand for the symmetric traceless part of a tensor. 
The short-distance coefficients $c_n$ are perturbatively calculable quantities
that do not depend on the meson state $|V\rangle$, 
while the long-distance matrix elements (LDMEs) of NRQCD operators between the 
vacuum $|0\rangle$
and the meson state $|V \rangle$ are nonperturbative quantities. 
We take the meson state $|V \rangle$ to be normalized nonrelativistically. 
In order to include the polarization vector in the short-distance 
coefficients $c_n$ in Eq.~(\ref{eq:Jpsifactorization}), we projected the NRQCD 
operators on the polarization vector of the state $\langle V |$.

In Eq.~(\ref{eq:Jpsifactorization}), we included operators that do not contain
the 
chromoelectric or chromomagnetic fields up to dimension 7, and operators that
do contain the  
chromoelectric or chromomagnetic fields up to dimension 6, 
all of which have definite total
angular momentum $J=1$, charge conjugation $C=-1$ and parity $P=-1$, which are
the same as $V = J/\psi$ or $\Upsilon(nS)$. 
Throughout this paper, 
we denote the operators that do not contain chromoelectric or
chromomagnetic fields as color-singlet operators, and the ones that do contain 
the chromoelectric or chromomagnetic fields as color-octet operators. 

Among all possible NRQCD operators, 
we included in Eq.~(\ref{eq:Jpsifactorization}) only the operators whose 
long-distance matrix elements (LDMEs) 
contribute to the amplitude up to relative order $v^4$, 
based on the conservative power counting of 
Refs.~\cite{Brambilla:2001xy, Brambilla:2002nu, Brambilla:2006ph,
Brambilla:2008zg}. 
In this power counting, the velocity scaling of the LDME of an NRQCD operator 
is determined by 
the dimension of the operator, where a power of $v$
is
associated to a unit of dimension, and by the contribution to the 
meson state of the $Q \bar Q$ Fock state created by the
operator. 
For $J/\psi$ or $\Upsilon(nS)$, the leading Fock state contains a $Q \bar Q$
in the color-singlet ${}^3S_1$ state, and this state has the scaling
$v^{-3/2}$. 
Subleading Fock states such as the ones that contain 
$Q \bar Q$ in a color-octet state, or the ones that contain $Q
\bar Q$ in a color-singlet $D$-wave state are suppressed by $v$ and $v^2$
compared to the leading Fock state, respectively. 
The color-singlet operator 
$\psi^\dag \bm{\sigma} \cdot \bm{\epsilon} (\lambda) \chi$ is the
lowest-dimensional operator that creates a $Q \bar Q$ in the leading Fock state
(${}^3S_1$), and so, the LDME 
$\langle V | \psi^\dag \bm{\sigma} \cdot \bm{\epsilon} (\lambda) \chi 
| 0 \rangle$ 
scales like $v^{3/2}$, and 
contributes to the amplitude at leading order in $v$. 
The LDMEs of the operators $\psi^\dag \bm{\sigma} \cdot \bm{\epsilon} (\lambda)
(-\tfrac{i}{2} \overleftrightarrow{\bm{D}})^2 \chi$ 
and $\psi^\dag \bm{\sigma} \cdot \bm{\epsilon} (\lambda)
(-\tfrac{i}{2} \overleftrightarrow{\bm{D}})^4 \chi$ 
scale like $v^{7/2}$ and $v^{11/2}$, respectively, 
because the operators create the $Q \bar Q$ in the leading Fock state and 
have dimensions that are higher than the lowest-dimensional operator by 2 and
4, respectively. Hence, these LDMEs contribute to the amplitude at
relative order $v^2$ and $v^4$, respectively. 
The operator $\psi^\dag \epsilon^i (\lambda) \sigma^j (-\tfrac{i}{2})^2 
\overleftrightarrow{\bm{D}}^{(i} \overleftrightarrow{\bm{D}}^{j)} \chi$
creates a color-singlet $Q \bar Q$ in a ${}^3D_1$ state. Since the 
$D$-wave Fock state has a contribution to the meson state suppressed by $v^2$ 
compared to the leading Fock state, the LDME 
$\langle V | \psi^\dag \epsilon^i (\lambda) \sigma^j (-\tfrac{i}{2})^2 
\overleftrightarrow{\bm{D}}^{(i} \overleftrightarrow{\bm{D}}^{j)} 
\chi | 0 \rangle$ scales like $v^{11/2}$ and 
contributes to the amplitude at relative order $v^4$. 
The color-octet operators in Eq.~(\ref{eq:Jpsifactorization}) 
create $Q \bar Q$ in color-octet states where either the orbital or the spin 
angular momentum is different from that of the leading Fock state by 1. 
The contributions of such Fock states are suppressed by $v$ compared to the
leading Fock state. Hence, the LDMEs 
$\langle V | \psi^\dag g_s \bm{B} \cdot \bm{\epsilon} (\lambda) \chi | 0
\rangle$,
$\langle V | \psi^\dag 
\bm{\sigma} \cdot \bm{\epsilon} (\lambda) 
\tfrac{1}{3} 
( \overleftrightarrow{\bm{D}} \cdot g_s \bm{E} 
+g_s \bm{E} \cdot \overleftrightarrow{\bm{D}} ) 
\chi | 0 \rangle$, 
and 
$\langle V | \psi^\dag \bm{\epsilon} (\lambda) \cdot 
\tfrac{1}{2} 
[\bm{\sigma} \times ( \overleftrightarrow{\bm{D}} \times g_s \bm{E}
- g_s \bm{E} \times \overleftrightarrow{\bm{D}} )] \chi | 0 \rangle$ 
scale like $v^{9/2}$, $v^{11/2}$, and $v^{11/2}$, respectively, and 
contribute to the amplitude at relative order $v^3$, $v^4$, and $v^4$, 
respectively. 
For later use, we also define ratios of LDMEs as follows:
\begin{subequations}
\begin{eqnarray}
\langle v^2_S \rangle_V &=&
\frac{1}{m^2} 
\frac{ \langle V | \psi^\dag \bm{\sigma} \cdot \bm{\epsilon} (\lambda)
(-\tfrac{i}{2} \overleftrightarrow{\bm{D}})^2 \chi | 0 \rangle }
{\langle V | \psi^\dag \bm{\sigma} \cdot \bm{\epsilon} (\lambda) \chi 
| 0 \rangle }, 
\\
\langle v^4_S \rangle_V &=& 
\frac{1}{m^4} 
\frac{ \langle V | \psi^\dag \bm{\sigma} \cdot
\bm{\epsilon} (\lambda)
(-\tfrac{i}{2} \overleftrightarrow{\bm{D}})^4 \chi | 0 \rangle}
{\langle V | \psi^\dag \bm{\sigma} \cdot \bm{\epsilon} (\lambda) \chi 
| 0 \rangle }, 
\\
\langle v^2_D \rangle_V &=& 
\frac{1}{m^2} 
\frac{\langle V | \psi^\dag 
\epsilon^i (\lambda) \sigma^j (-\tfrac{i}{2})^2 
\overleftrightarrow{\bm{D}}^{(i} \overleftrightarrow{\bm{D}}^{j)} 
\chi | 0 \rangle}
{\langle V | \psi^\dag \bm{\sigma} \cdot \bm{\epsilon} (\lambda) \chi 
| 0 \rangle }, \\
\langle B \rangle_V &=&
\frac{1}{m^2} \frac{\langle V | \psi^\dag g_s \bm{B} \cdot 
\bm{\epsilon} (\lambda)  \chi | 0 \rangle}
{\langle V | \psi^\dag \bm{\sigma} \cdot \bm{\epsilon} (\lambda) \chi 
| 0 \rangle }, \\
\langle DE_0 \rangle_V &=&
\frac{1}{m^3} \frac{\langle V | \psi^\dag 
\bm{\sigma} \cdot \bm{\epsilon} (\lambda) 
\tfrac{1}{3} 
( \overleftrightarrow{\bm{D}} \cdot g_s \bm{E} 
+g_s \bm{E} \cdot \overleftrightarrow{\bm{D}} ) 
\chi | 0 \rangle}
{\langle V | \psi^\dag \bm{\sigma} \cdot \bm{\epsilon} (\lambda) \chi 
| 0 \rangle }, \\
\langle DE_1 \rangle_V &=&
\frac{1}{m^3} 
\frac{\langle V | \psi^\dag \bm{\epsilon} (\lambda)  \cdot 
\tfrac{1}{2} 
[\bm{\sigma} \times ( \overleftrightarrow{\bm{D}} \times g_s \bm{E}
- g_s \bm{E} \times \overleftrightarrow{\bm{D}} )] \chi | 0 \rangle}
{\langle V | \psi^\dag \bm{\sigma} \cdot \bm{\epsilon} (\lambda) \chi 
| 0 \rangle }. 
\end{eqnarray}
\end{subequations}

There is a color-singlet operator of dimension 7 that does not appear in 
Eq.~(\ref{eq:Jpsifactorization}), which is given by 
$\frac{1}{2} \psi^\dag \epsilon^i (\lambda) \sigma^j (-\tfrac{i}{2})^2 
\{ \overleftrightarrow{\bm{D}}^{(i} \overleftrightarrow{\bm{D}}^{j)} , 
(-\tfrac{i}{2} \overleftrightarrow{\bm{D}})^2 \} \chi$.
Because this operator creates a $Q \bar Q$ in a 
${}^3D_1$ state, its LDME scales like $v^{15/2}$ and contributes to the amplitude at relative order
$v^6$. Similarly, the color-octet operators of dimension 6 given by 
$\psi^\dag \bm{\epsilon} (\lambda) \cdot 
\frac{i}{2} ( \overleftrightarrow{\bm{D}} \times g_s \bm{E}
+ g_s \bm{E} \times \overleftrightarrow{\bm{D}} ) \chi$
and 
$\psi^\dag \epsilon^i (\lambda) \sigma^j 
( \overleftrightarrow{\bm{D}}^{(i} g_s \bm{E}^{j)} 
+ g_s \bm{E}^{(i} \overleftrightarrow{\bm{D}}^{j)} ) \chi$ 
do not appear in Eq.~(\ref{eq:Jpsifactorization}) because 
their LDMEs contribute to the amplitude at relative order $v^5$ and $v^6$,
respectively. The velocity scalings of these LDMEs can also 
be determined from the
Gremm--Kapustin relations in Eqs.~(\ref{eq:Gremm3}) and (\ref{eq:Gremm4}). 

If we follow the power counting of Ref.~\cite{Bodwin:1994jh}, 
the color-octet LDMEs except for 
$\langle V | \psi^\dag \tfrac{1}{3} (\overleftrightarrow{\bm{D}} \cdot g_s 
\bm{E} 
+g_s \bm{E} \cdot \overleftrightarrow{\bm{D}} ) \bm{\sigma} \cdot \bm{\epsilon}
(\lambda)  \chi | 0 \rangle $
are suppressed beyond relative order $v^4$ and do not appear at the current
level of accuracy. 

We compute the short-distance coefficients $c_n$ appearing in 
Eq.~(\ref{eq:Jpsifactorization}) at leading order in $\alpha_s$ 
by using the perturbative matching conditions obtained by replacing the meson 
state $V$ with a perturbative $Q \bar Q$ or a $Q \bar Q g$ state. Since the
expression in Eq.~(\ref{eq:Jpsifactorization}) is only valid to a limited
accuracy in $v$, we expand the perturbative amplitude in powers of the 
$3$-momenta of the $Q$, $\bar Q$, and the gluon, and truncate the series to
the
desired accuracy. 
We follow a method used in Refs.~\cite{Braaten:1996rp, 
Brambilla:2017kgw}, 
that consists in not projecting to a 
specific color, spin or orbital angular momentum of the $Q \bar Q$
state, but  
instead, in only requiring the $Q \bar Q$ or the $Q \bar Q g$ state 
to have the same $J^{PC}= 1^{--}$ as the meson state $V$. 
This method has the advantage that fewer
matching conditions are required to compute the short-distance coefficients. 
The caveat is that specific expressions for the matching conditions can be more 
complicated than when projected to specific color, spin, and orbital angular 
momentum states. 
Therefore, this method is suitable for computer-aided, automatized 
calculations. 
Also, this method can require including NRQCD operators that have the same 
dimensions as the ones appearing in Eq.~(\ref{eq:Jpsifactorization}) but have
LDMEs that are suppressed beyond relative order $v^4$, 
because the matching conditions obtained in this way
do not depend on the probabilities of the $Q \bar Q$ Fock states to be found 
in the meson state.
Hence, in the calculation of the
short-distance coefficients, we include all color-singlet operators of
dimensions up to 7, and color-octet operators of dimensions up to 6, 
that have $J^{PC}=1^{--}$. 

If we replace the meson state $V$ with a perturbative $Q \bar Q$ state, 
the amplitude occurs from order $g_s^0$, and the color-octet operators 
do not contribute to the amplitude at this order. 
We include all color-singlet operators up to dimension 7, which 
contain at most 4 covariant derivatives. Hence, we must consider the 
production amplitude of a $Q \bar Q$ state at up to
fourth order in the relative momentum of the $Q$ and the $\bar Q$.
We use the kinematical configuration given in Appendix~\ref{sec:kinematics} 
where the relative 3-momentum between the $Q$ and the $\bar Q$ is given by
$\bm{q}$. The production amplitude of a $Q \bar Q$ state with 
$J^{PC} = 1^{--}$ and a photon is given at order $g_s^0$ by 
\begin{eqnarray}
\label{eq:QQfactorization}%
&& \hspace{-5ex}
i {\cal M} [H \to Q \bar Q(J^{PC}=1^{--}) + \gamma] \nonumber  \\
&=& 
c_0 \langle Q \bar Q | \psi^\dag \bm{\sigma} \cdot \bm{\epsilon} (\lambda) \chi | 0 \rangle 
+ \frac{c_{\bm{D}^2}}{m^2} \langle Q \bar Q | \psi^\dag \bm{\sigma} \cdot \bm{\epsilon} (\lambda)
(-\tfrac{i}{2} \overleftrightarrow{\bm{D}})^2 \chi | 0 \rangle 
\nonumber \\ && 
+ \frac{c_{\bm{D}^{(i} \bm{D}^{j)}}}{m^2} \langle Q \bar Q | \psi^\dag 
\epsilon^i (\lambda)
\sigma^j
(-\tfrac{i}{2})^2 
\overleftrightarrow{\bm{D}}^{(i} 
\overleftrightarrow{\bm{D}}^{j)} 
\chi | 0 \rangle 
\nonumber \\ && 
+ \frac{c_{\bm{D}^4}}{m^4} \langle Q \bar Q | \psi^\dag \bm{\sigma} \cdot \bm{\epsilon} (\lambda)
(-\tfrac{i}{2} \overleftrightarrow{\bm{D}})^4 \chi | 0 \rangle 
\nonumber \\ && 
+ \frac{c_{\bm{D}^2 \bm{D}^{(i} \bm{D}^{j)}}}{m^4} 
\langle Q \bar Q | \tfrac{1}{2} \psi^\dag \epsilon^i (\lambda) \sigma^j 
(-\tfrac{i}{2})^2 
\{ \overleftrightarrow{\bm{D}}^{(i} \overleftrightarrow{\bm{D}}^{j)} ,
(-\tfrac{i}{2} \overleftrightarrow{\bm{D}} )^2 \}
\chi | 0 \rangle 
 + O(g_s, (|\bm{q}|/m)^5), 
\end{eqnarray}
where we have included all color-singlet operators with $J^{PC} = 1^{--}$ up to
dimension 7. 
We take the states $|Q\rangle$ and the $|\bar Q\rangle$ to be nonrelativistically normalized. 
We determine the short-distance coefficients $c_0$, 
$c_{\bm{D}^2}$, $c_{\bm{D}^{(i} \bm{D}^{j)}}$, and $c_{\bm{D}^4}$, along with 
$c_{\bm{D}^2 \bm{D}^{(i} \bm{D}^{j)}}$ that does not appear in
Eq.~(\ref{eq:Jpsifactorization}), by computing the
left- and right-hand sides in perturbative QCD and 
perturbative NRQCD, respectively, and comparing the two sides order by
order in the expansion in powers of $\bm{q}$ up to fourth order. 

In order to compute the remaining short-distance coefficients corresponding to
the color-octet LDMEs, we consider the production amplitude of a 
$Q \bar Q g$ state with $J^{PC} = 1^{--}$ which occurs from order $g_s$. 
We use the kinematical configuration given in Appendix~\ref{sec:kinematics} 
where the relative 3-momentum between the $Q$ and the $\bar Q$ is given by
$\bm{q}_1$ and the relative 3-momentum between the $Q \bar Q$ pair and the gluon is
given by $\bm{q}_2$. 
At order $g_s$, the color-octet LDMEs that appear in 
Eq.~(\ref{eq:Jpsifactorization}) have matrix elements that are 
either linear or quadratic in the momenta $\bm{q}_1$ or $\bm{q}_2$ when the
meson state $V$ is replaced by the $Q \bar Q g$ state.
We must also include all color-octet operators of dimensions up to 6 with 
$J^{PC} = 1^{--}$ that do not appear in Eq.~(\ref{eq:Jpsifactorization}), 
whose matrix elements can also be either linear or quadratic in the momenta 
$\bm{q}_1$ or $\bm{q}_2$. 
The color-singlet operators in Eq.~(\ref{eq:Jpsifactorization}) can also 
contribute to the $Q \bar Q g$ amplitude at order $g_s$ 
through the gauge fields in the
covariant derivatives and through insertions of NRQCD vertices. 
An NRQCD vertex insertion at order $g_s$ involves a heavy-quark propagator,
which can produce a factor of $1/|\bm{q}_2|$. 
Hence, it is necessary to include all color-singlet operators that contain at 
most 3 covariant derivatives. 
Since the lowest-dimensional color-singlet operator we consider is
of dimension 3, and the highest-dimensional color-octet operators are of
dimension 6, we need to include NRQCD vertices up to
$1/m^3$ accuracy. That is, we need to consider two-fermion operators of 
dimensions up to 7 in the NRQCD Lagrangian, which are given
by~\cite{Manohar:1997qy} 
\begin{eqnarray}
\label{eq:lagrangian}
{\cal L}_{\rm 2-f} 
&=& \psi^\dag \bigg( i D_0 + \frac{\bm{D}^2}{2 m} 
+ \frac{\bm{\sigma} \cdot g_s \bm{B}}{2 m} 
+ \frac{(\bm{D} \cdot g_s \bm{E})}{2 m} 
- \frac{\bm{\sigma} \cdot [-i \bm{D} \times , g_s \bm{E}]}{8 m^2}
\nonumber \\ && \hspace{5ex}
+ \frac{\bm{D}^4}{8 m^3} + \frac{\{\bm{D}^2, \bm{\sigma} \cdot g_s \bm{B}\}}{8
m^3} + \ldots \bigg) \psi
+ \textrm{c.c.}, 
\end{eqnarray}
where c.c.\ stands for the charge-conjugated contribution of the preceding
terms. 
Since we only consider the matching at tree level, we only include the
Wilson coefficients at order $\alpha_s^0$ in Eq.~(\ref{eq:lagrangian}). 
The production amplitude of a $Q \bar Q g$ state 
with $J^{PC} = 1^{--}$ and a photon at order $g_s$ is given by 
\begin{eqnarray}
\label{eq:QQgfactorization}%
&& \hspace{-5ex}
i {\cal M} [H \to Q \bar Q g (J^{PC}=1^{--}) + \gamma ] 
\nonumber \\ 
&=& 
c_0 \langle Q \bar Q g | \psi^\dag \bm{\sigma} \cdot \bm{\epsilon} (\lambda) \chi | 0 \rangle 
+ \frac{c_{\bm{D}^2}}{m^2} \langle Q \bar Q g | \psi^\dag \bm{\sigma} \cdot \bm{\epsilon} (\lambda)
(-\tfrac{i}{2} \overleftrightarrow{\bm{D}})^2 \chi | 0 \rangle 
\nonumber \\ && 
+ \frac{c_{\bm{D}^{(i} \bm{D}^{j)}}}{m^2} \langle Q \bar Q g | \psi^\dag 
\epsilon^i (\lambda)
\sigma^j
(-\tfrac{i}{2})^2 
\overleftrightarrow{\bm{D}}^{(i} 
\overleftrightarrow{\bm{D}}^{j)} 
\chi | 0 \rangle 
\nonumber \\ && 
+ \frac{c_B}{m^2} \langle Q \bar Q g | 
\psi^\dag g_s \bm{B} \cdot \bm{\epsilon} (\lambda) \chi | 0 \rangle 
\nonumber\\ && 
+ \frac{c_{DE_0}}{m^3} 
\langle Q \bar Q g | \psi^\dag 
\bm{\epsilon} (\lambda) \cdot \bm{\sigma}
\tfrac{1}{3} 
( \overleftrightarrow{\bm{D}} \cdot g_s \bm{E} 
+g_s \bm{E} \cdot \overleftrightarrow{\bm{D}} ) \chi | 0 \rangle 
\nonumber\\ && 
+ \frac{c_{DE_1}}{m^3} 
\langle Q \bar Q g | \psi^\dag 
\bm{\epsilon} (\lambda) \cdot 
\tfrac{1}{2} 
[\bm{\sigma} \times ( \overleftrightarrow{\bm{D}} \times g_s \bm{E} 
- g_s \bm{E} \times \overleftrightarrow{\bm{D}} )] \chi | 0 \rangle 
\nonumber\\ && 
+ \frac{c_{DE_1'}}{m^3} 
\langle Q \bar Q g | \psi^\dag 
\bm{\epsilon} (\lambda) \cdot 
\tfrac{i}{2} 
( \overleftrightarrow{\bm{D}} \times g_s \bm{E} 
+g_s \bm{E} \times \overleftrightarrow{\bm{D}} ) \chi | 0 \rangle 
\nonumber\\ && 
+ \frac{c_{DE_2}}{m^3} 
\langle Q \bar Q g | \psi^\dag 
\epsilon^i (\lambda) \sigma^j 
( \overleftrightarrow{\bm{D}}^{(i} g_s \bm{E}^{j)} 
+ g_s \bm{E}^{(i} \overleftrightarrow{\bm{D}}^{j)} )
\chi | 0 \rangle 
+ O(g_s^2, |\bm{q}_i|^3/m^3),
\end{eqnarray}
where the left-hand side is calculated in perturbative QCD and is expanded in
powers of $\bm{q}_1$ and $\bm{q}_2$ up to quadratic accuracy. 
We again take the states $|Q\rangle$ and $|\bar Q\rangle$ to be nonrelativistically normalized. 
We determine the short-distance coefficients 
$c_{B}$, $c_{DE_0}$, and $c_{DE_1}$, along with 
$c_{DE_1'}$ and $c_{DE_2}$ that do not appear in
Eq.~(\ref{eq:Jpsifactorization}) by computing the left- and right-hand sides in
perturbative QCD and NRQCD, respectively, and comparing the two sides order by
order in the expansion in powers of $\bm{q}_1$ and $\bm{q}_2$ up to quadratic
order.\footnote{The operator 
$\psi^\dag \bm{\epsilon} (\lambda) \cdot \tfrac{i}{2} 
( \overleftrightarrow{\bm{D}} \times g_s \bm{E} 
+g_s \bm{E} \times \overleftrightarrow{\bm{D}} ) \chi$ does not appear in
Ref.~\cite{Brambilla:2006ph}. 
If we compute the NRQCD matrix elements on the right-hand side
of Eq.~(\ref{eq:QQgfactorization}) explicitly, 
the matrix element of this operator is the only 
matrix element that is quadratic in $\bm{q}_2$. 
Hence, if we ignore the contribution that is proportional to $|\bm{q}_2|^2$
on both sides of Eq.~(\ref{eq:QQgfactorization}), we can ignore this operator 
from the matching condition without affecting the calculation of the 
short-distance coefficients in Eq.~(\ref{eq:Jpsifactorization}). 
} 

In the following sections, we compute the short-distance coefficients $c_n$
explicitly. We first calculate the $c_n$ in fixed-order perturbation theory,
where the QCD amplitudes on the left-hand sides of 
Eqs.~(\ref{eq:QQfactorization}) and (\ref{eq:QQgfactorization}) are computed at 
leading order in $\alpha_s$. We obtain corrections to the direct amplitude 
of relative order $v^4$ which is new in this work, and
reproduce the known order $v^2$ correction in the fixed-order calculation. 
We then compute the $c_n$ in the light-cone approach, 
which is valid at leading order in $m^2/m_H^2$, 
that allows us to resum logarithms of $m_H^2/m^2$ to all orders in $\alpha_s$. 
We obtain new corrections of relative order $v^4$ in the light-cone 
approach, and reproduce the previously calculated order $v^2$ correction. 
We include the order $\alpha_s$ correction 
to the direct amplitude using the light-cone approach.

\subsection{Fixed-order calculation}
\label{sec:fixedorder}%

At order $g_s^0$, the direct amplitude for 
$H \to Q \bar Q + \gamma$ is given by 
\begin{eqnarray}
\label{eq:QQamplitude}%
&& \hspace{-5ex} 
i {\cal M}_{\rm dir} (H \to Q \bar Q + \gamma) 
\nonumber \\
&=& 
-i e e_Q y_Q 
\bar u (p_1) 
\bigg[
\frac{(-p\!\!\!/_2 -p\!\!\!/_\gamma + m) \epsilon\!\!\!/^*_\gamma}
{ (p_2+p_\gamma)^2 -m^2+i \varepsilon} 
+ 
\frac{\epsilon\!\!\!/^*_\gamma (p\!\!\!/_1 +p\!\!\!/_\gamma + m)}
{ (p_1+p_\gamma)^2 -m^2+i \varepsilon} 
\bigg] v(p_2) ,
\end{eqnarray}
where $p_\gamma$ and $\epsilon_\gamma^*$ are the momentum and the polarization
vector for the photon in the final state. 
We use the physical gauge for the photon polarization vector, so that 
$\epsilon_\gamma^* \cdot p_\gamma = 0$.
Here,
$e = \sqrt{4 \pi \alpha}$ is the electric charge, $e_Q$ is the fractional
charge of the heavy quark $Q$, and
 $y_Q= \overline{m} (\mu) (\sqrt{2} G_F)^{\frac{1}{2}}$
is the Yukawa coupling of the Higgs boson and $Q$, with $G_F$ the Fermi constant.
The momenta of the $Q$ and $\bar Q$ are given by $p_1$ and $p_2$, 
respectively, so that the momentum of the $H$ is $P_H = p_1 + p_2 + p_\gamma$. 
This implies $m_H^2 = (p_1 + p_2 + p_\gamma)^2
= 2 p_\gamma \cdot (p_1 + p_2) + (p_1 + p_2)^2$, 
so that in the rest frame of the $Q \bar Q$, 
\begin{equation}
p_\gamma^0 = |\bm{p}_\gamma| = \frac{m_H^2 - (p_1+p_2)^2}{2 \sqrt{(p_1+p_2)^2}}
= \frac{m_H^2 - 4 m^2 -4 \bm{q}^2}{4 \sqrt{m^2 + \bm{q}^2}},
\end{equation}
where $q = \frac{1}{2} (p_1 - p_2)$. 
We choose the heavy-quark mass 
appearing in $y_Q$ to be $\overline{m} (\mu)$, which is 
the $\overline{\rm MS}$ mass of the heavy quark $Q$ at scale $\mu$; as we
will see in the next section, this choice simplifies the logarithms that appear 
in the order $\alpha_s$ correction. 
Since $C$, $P$, and $T$ are conserved in the amplitude in
Eq.~(\ref{eq:QQamplitude}), 
the $Q \bar Q$ can only be created with $C=-1$. 
We first express the Dirac bilinears $\bar u (p_1) \gamma^\mu v(p_2)$ and 
$\bar u (p_1) \gamma^\mu \gamma^\nu v(p_2)$ in terms of the 3-momenta of the
$Q$ and $\bar Q$ in the $Q \bar Q$ rest frame. This can be accomplished by
using the method described in Appendix~\ref{sec:spinprj}. 
Then, we expand the amplitude in powers
of $\bm{q}$, keeping terms up to relative order $\bm{q}^4/m^4$. 
The resulting expression for the amplitude is then a linear combination of the
Cartesian tensors built from $\xi^\dag \bm{\sigma} \eta$ and $\bm{q}$ 
of the form $\xi^\dag \sigma^i \eta q^j \cdots q^k$ up to rank 5 
and of the form $\xi ^\dag \eta q^i q^j \cdots q^k$ up to rank 4. 
The contribution from these Cartesian tensors to the total angular momentum $J=1$ 
can be obtained by 
a reduction method developed in Ref.~\cite{CoopeSnider}.
Finally, the $P=-1$ contribution is obtained by keeping only the 
contribution odd in parity, where the parity transform of the $Q \bar Q$
amplitude is given by the replacements 
$\bm{q} \to - \bm{q}$, 
$\xi^\dag \bm{\sigma} \eta \to -\xi^\dag \bm{\sigma} \eta$, and 
$\xi^\dag \eta \to -\xi^\dag \eta$. 
We use the {\it Mathematica} package {\scshape FeynCalc}~\cite{Mertig:1990an,
Shtabovenko:2016sxi} 
and the {\scshape FeynOnium}~\cite{FeynOnium} package to 
automatize the calculation of the amplitude and the consequent reduction to the
$J^{PC} =1^{--}$ contribution. 
By comparing the $J^{PC} = 1^{--}$ contribution of the $Q \bar Q$ amplitude 
with the right-hand side of 
Eq.~(\ref{eq:QQfactorization}), we obtain the short-distance coefficients 
\begin{subequations}
\label{eq:WC_singlet}%
\begin{eqnarray}
c_0 &=& - i \frac{e e_Q y_Q}{m} 
\bm{\epsilon}_\gamma^* \cdot \bm{\epsilon}^* (\lambda), \\
c_{\bm{D}^2} &=& i \frac{e e_Q y_Q}{m} 
\frac{3 - 7r}{6 (1-r)} 
\bm{\epsilon}_\gamma^* \cdot \bm{\epsilon}^* (\lambda), \\
c_{\bm{D}^{(i} \bm{D}^{j)}} &=& - i \frac{e e_Q y_Q}{m} 
\frac{3 +17 r}{10 (1-r)} 
\bm{\epsilon}_\gamma^* \cdot \bm{\epsilon}^* (\lambda), \\
c_{\bm{D}^4} &=& - i \frac{e e_Q y_Q}{m} 
\frac{43 - 110 r+147 r^2}{120 (1 - r)^2} 
\bm{\epsilon}_\gamma^* \cdot \bm{\epsilon}^* (\lambda), \\
c_{\bm{D}^2 \bm{D}^{(i} \bm{D}^{j)}} &=& 
i \frac{e e_Q y_Q}{m} 
\frac{83  + 2 r -645 r^2}{280 (1-r)^2} 
\bm{\epsilon}_\gamma^* \cdot \bm{\epsilon}^* (\lambda), 
\end{eqnarray}
\end{subequations}
where we define $r \equiv \frac{4 m^2}{m_H^2}$. 
The short-distance coefficients $c_0$ and $c_{\bm{D}^2}$ agree with
Refs.~\cite{Bodwin:2013gca, Bodwin:2014bpa}, except that our results differ 
by an overall sign that originates from the sign convention of the $J=1$ state
employed in Refs.~\cite{Bodwin:2013gca, Bodwin:2014bpa}. 
We also reproduce the short-distance coefficient $c_{\bm{D}^4}$ 
that can be obtained from Ref.~\cite{Bodwin:2014bpa}. 
The results for $c_{\bm{D}^{(i} \bm{D}^{j)}}$
and $c_{\bm{D}^2 \bm{D}^{(i} \bm{D}^{j)}}$ are new. 

The remaining short-distance coefficients 
corresponding to the color-octet LDMEs 
are computed from the direct amplitude for 
$H \to Q \bar Q g + \gamma$ at order $g_s$, which is given by
\begin{eqnarray}
\label{eq:QQgamplitude}%
i {\cal M}_{\rm dir} (H \to Q \bar Q g + \gamma) 
&=& 
-i g_s e e_Q y_Q 
\bar u (p_1) T^a
\nonumber \\ && \times 
\bigg\{
\frac{\epsilon\!\!\!/^*_g (p\!\!\!/_1+k\!\!\!/_g + m) 
(-p\!\!\!/_2 - p\!\!\!/_\gamma + m) \epsilon\!\!\!/^*_\gamma}
{[(p_1 +k_g)^2 -m^2 +i \varepsilon] 
[(p_2 +p_\gamma)^2 -m^2 +i \varepsilon] }
\nonumber \\ && + 
\frac{(-p\!\!\!/_2 -k\!\!\!/_g -p\!\!\!/_\gamma + m) 
\epsilon\!\!\!/^*_\gamma (-p\!\!\!/_2-k\!\!\!/_g+m) 
\epsilon\!\!\!/^*_g}
{[ (p_2+k_g+p_\gamma)^2-m^2+i\varepsilon]
[ (p_2+k_g)^2-m^2+i\varepsilon]}
\nonumber \\ && + 
\frac{ (-p\!\!\!/_2-k\!\!\!/_g-p\!\!\!/_\gamma + m) 
\epsilon\!\!\!/^*_g
(-p\!\!\!/_2-p\!\!\!/_\gamma+m)
\epsilon\!\!\!/^*_\gamma }
{[ (p_2+k_g+p_\gamma)^2-m^2+i \varepsilon]
[ (p_2+p_\gamma)^2-m^2+i \varepsilon]}
\nonumber \\ && + 
\frac{
\epsilon\!\!\!/^*_\gamma
(p\!\!\!/_1+p\!\!\!/_\gamma+m) 
(-p\!\!\!/_2-k\!\!\!/_g+m)
\epsilon\!\!\!/^*_g 
}
{[ (p_2+k_g)^2 -m^2+i \varepsilon]
[ (p_1+p_\gamma)^2-m^2+i \varepsilon]}
\nonumber \\ && + 
\frac{
\epsilon\!\!\!/^*_g
(p\!\!\!/_1+k\!\!\!/_g+m)
\epsilon\!\!\!/^*_\gamma
(p\!\!\!/_1+k\!\!\!/_g+p\!\!\!/_\gamma+m)
}
{[ (p_1+k_g+p_\gamma)^2 -m^2+i \varepsilon]
[ (p_1+k_g)^2 - m^2 +i \varepsilon]}
\nonumber \\ && + 
\frac{
\epsilon\!\!\!/^*_\gamma
(p\!\!\!/_1+p\!\!\!/_\gamma + m)
\epsilon\!\!\!/^*_g
(p\!\!\!/_1+p\!\!\!/_\gamma + k\!\!\!/_g + m)
}{[ (p_1+p_\gamma)^2 -m^2+i \varepsilon]
[ (p_1+p_\gamma + k_g)^2  -m^2+i \varepsilon]}
\bigg\} v(p_2), 
\end{eqnarray}
where $\epsilon_g^*$ is the polarization vector of the gluon. 
The total momentum $P$ of the $Q \bar Q g$ state is given by 
$P = p_1 + p_2 + k_g$, and the momentum of the $H$ is given by 
$P_H = P + p_\gamma$, 
so that in the rest frame of the $Q \bar Q g$,
\begin{equation}
p_\gamma^0 = |\bm{p}_\gamma| = \frac{m_H^2 - P_0^2}{2 P_0},
\end{equation}
where $P_0 = 2 |\bm{q}_2| + \sqrt{(\bm{q}_1 + \bm{q}_2)^2 + m^2} 
+ \sqrt{(\bm{q}_1 - \bm{q}_2)^2 + m^2}$, 
$q_1 = \frac{1}{2} (p_1 - p_2)$,
and $q_2 = \frac{1}{6} (2 k_g - p_1 -p_2)$. 
Due to $C$ conservation  at this order, the $Q \bar Q g$ can only
be produced in a color-singlet $C=-1$ state. 
We use again the {\it Mathematica} package {\scshape FeynCalc}
and the {\scshape FeynOnium} package to 
automatize the calculation of the $Q \bar Q g$ amplitude. 
After expressing the amplitude in terms of the 3-momenta of the
$Q$, $\bar Q$ and $g$, we expand the amplitude in powers of $\bm{q}_1$ and
$\bm{q}_2$ up to relative order 
$\bm{q}_1^2 / m^2$,
$\bm{q}_2^2 / m^2$,
and 
$|\bm{q}_1| |\bm{q}_2| / m^2$ 
The resulting expression for the amplitude is then a linear combination of the
Cartesian tensors built from 
$\xi^\dag \bm{\sigma} \eta$, 
$\bm{\epsilon}_g^*$, $\bm{q}_1$, and $\bm{q}_2$ up to rank 4. 
In order to keep only the contribution with negative parity, 
we keep only the contribution odd in parity, 
where the parity transform of the $Q \bar Q g$
amplitude is given by the replacements 
$\bm{q}_1 \to - \bm{q}_1$, 
$\bm{q}_2 \to - \bm{q}_2$, 
$\bm{\epsilon}_g^* \to - \bm{\epsilon}_g^*$, 
$\xi^\dag \bm{\sigma} \eta \to -\xi^\dag \bm{\sigma} \eta$, and 
$\xi^\dag \eta \to -\xi^\dag \eta$. 
After reducing the Cartesian tensors of odd rank to total angular momentum
$J=1$, we compare the $J^{PC}=1^{--}$ contribution with the amplitude 
on the right-hand side of 
Eq.~(\ref{eq:QQgfactorization}) to obtain the short-distance coefficients 
\begin{subequations}
\label{eq:WC_octet}%
\begin{eqnarray}
c_{B} &=& -i \frac{e e_Q y_Q}{m} 
\bm{\epsilon}_\gamma^* \cdot \bm{\epsilon}^* (\lambda), \\
c_{DE_0} &=& i \frac{e e_Q y_Q}{m} 
\frac{3 -6 r +5 r^2}{4 (1-r)^2} 
\bm{\epsilon}_\gamma^* \cdot \bm{\epsilon}^* (\lambda), \\
c_{DE_1} &=& i \frac{e e_Q y_Q}{m} 
\frac{3 -4 r + 5 r^2}{8 (1-r)^2} 
\bm{\epsilon}_\gamma^* \cdot \bm{\epsilon}^* (\lambda), \\
c_{DE_1'} &=& i \frac{e e_Q y_Q}{m} 
\frac{5 -2 r}{4 (1-r)} 
\bm{\epsilon}_\gamma^* \cdot \bm{\epsilon}^* (\lambda), \\
c_{DE_2} &=& -i \frac{e e_Q y_Q}{m} 
\frac{(3-5 r) (3 +7 r)}{40 (1 - r)^2} 
\bm{\epsilon}_\gamma^* \cdot \bm{\epsilon}^* (\lambda), 
\end{eqnarray}
\end{subequations}
where $r = \frac{4 m^2}{m_H^2}$. 

The short-distance coefficients in Eqs.~(\ref{eq:WC_singlet}) and
(\ref{eq:WC_octet}) allow us to compute the direct amplitude to relative order
$v^4$ accuracy. 
Because the indirect amplitude will be available only at
leading order in $m^2/m_H^2$ (see Sec.~\ref{sec:indirect}), 
only the limit $m^2/m_H^2 \to 0$ of the short-distance coefficients in 
Eqs.~(\ref{eq:WC_singlet}) and (\ref{eq:WC_octet}) 
will be employed in the total decay amplitude. 
This amounts to setting $r=0$.

We note that the method described in this section can be easily applied to
compute contributions of the QCD amplitude with different $J^{PC}$. For
example, the $J^{PC} = 1^{+-}$ contribution can be 
obtained by keeping parity-even terms in the amplitude. The $J^{PC} =
1^{+-}$ contribution can then be
used to obtain the 
short-distance
coefficients for the $H \to h_c + \gamma$ amplitude. We show this result in
Appendix~\ref{sec:axialvector}. 

\subsection{Light-cone calculation}
\label{sec:lightcone}

In the fixed-order calculation, the short-distance coefficients contain
contributions from the scales $m$ and $m_H$. Since $m_H$ is much larger than
$m$, the logarithms of $m_H^2/m^2$ that appear in corrections of higher orders 
in $\alpha_s$ can potentially spoil the convergence of the perturbation series. 
If we work at leading power in $m^2/m_H^2$, the light-cone approach provides a
factorization formula that separates the contribution at the scale $m$ from
the contribution at the scale $m_H$~\cite{Lepage:1980fj, Chernyak:1983ej}. 
The light-cone approach also enables us to resum the logarithms of $m_H^2/m^2$ 
to all orders in $\alpha_s$ by solving an evolution equation. 
The factorization formula is given by~\cite{Lepage:1980fj, Chernyak:1983ej, 
Wang:2013ywc}
\begin{equation}
\label{eq:LCDAfac}%
i {\cal M}_{\rm dir} (H \to V + \gamma)  = 
\frac{i}{2} e e_Q y_Q
\epsilon_\gamma^*{} \cdot \epsilon^* (\lambda)
f_{V}^\perp (\mu) 
\int_0^1 dx\, T_H (x,\mu) 
\phi_{V}^\perp (x,\mu) +O(m^2/m_H^2), 
\end{equation}
where $T_H (x,\mu)$ is the perturbative hard part that contains the 
contribution at the scale
$m_H$, while the contribution at scales $m$ and below are contained in
the decay constant $f_{V}^\perp (\mu)$ and the light-cone distribution 
amplitude (LCDA) $\phi_{V}^\perp (x,\mu)$ of the meson $V$. 
The decay constant and the LCDA are nonperturbative quantities defined by 
\begin{equation}
\label{eq:LCDAdef}%
f_V^\perp (\mu)
\epsilon_\perp^*{}^\alpha (\lambda)
\phi_{V}^\perp (x,\mu)
=
\langle V | {\cal Q}^\alpha (x) | 0 \rangle.
\end{equation}
 The nonlocal operator ${\cal Q}^\alpha (x)$ is defined by 
\begin{equation}
{\cal Q}^\alpha (x) = 
\int \frac{d \omega}{2 \pi} 
e^{-i (x-1/2) \omega \bar n \cdot P} 
(\bar Q W_c) (\omega \bar n/2) \bar n\!\!\!/ \gamma_\perp^\alpha
(W_c^\dag Q) (-\omega \bar n/2),
\end{equation}
where $P$ is the momentum of the meson $V$, 
and the decay constant $f_V^\perp(\mu)$ is defined by
integrating Eq.~(\ref{eq:LCDAdef}) over $x$ and considering 
the normalization of the LCDA, which is given by 
$\int_0^1 dx \, \phi_{V}^\perp (x, \mu) = 1$.
Here, $Q(x)$ is the QCD quark field. 
The light-cone vectors $n$ and $\bar n$ are given by 
$\bar n = \frac{m_H}{p_\gamma \cdot P_H} p_\gamma$
and $n = \frac{2}{m_H} P_H - \bar n$, which satisfy $n^2 = \bar n^2 = 0$
and $\bar n \cdot n = 2$. Here, $P_H$ and $p_\gamma$ are the momenta of the $H$
and the photon, respectively. 
 For any 4-vector $a^\mu$, we define $a^\mu_\perp\equiv a^\mu-\frac{n^\mu}{2}\bar{n}\cdot a-\frac{\bar{n}^\mu}{2}{n}\cdot a$.
The Wilson line 
$W_c(x)={\cal P} \exp \left[-i g \int_{-\infty}^0 ds \, \bar n \cdot A(x+s \bar n)
\right]$, where ${\cal P}$ is the path-ordering operator, ensures the gauge invariance of the nonlocal operator.  
The hard part $T_H (x,\mu)$ has been computed to next-to-leading order (NLO) 
accuracy in $\alpha_s$ 
and is given by~\cite{Wang:2013ywc, Koenig:2015pha}
\begin{eqnarray}
\label{eq:TH}%
T_H (x,\mu) &=& \frac{1}{x (1-x)}
+ \frac{\alpha_s (\mu) C_F}{4 \pi} 
\frac{1}{x (1-x)}
\nonumber \\ && 
\times \bigg[
2 \left( \log \frac{m_H^2}{\mu^2} -i \pi \right) 
\log x (1-x) + \log^2 x+\log^2 (1-x) -3 \bigg]
+ O(\alpha_s^2),
\end{eqnarray}
where $C_F = \frac{N_c^2 - 1}{2 N_c}$, and $N_c = 3$ is the number of
colors. The imaginary part in Eq.~(\ref{eq:TH}) comes from the order $\alpha_s$
correction where the virtual lines can be on shell.
The expression for the NLO correction to $T_H(x,\mu)$ in Eq.~(\ref{eq:TH}) is 
valid when 
the decay constant $f_V^\perp (\mu)$ and the LCDA $\phi_V^\perp (x,\mu)$ are
renormalized in the $\overline{\rm MS}$ scheme, and 
the heavy-quark mass in the Yukawa coupling $y_Q$ is the 
$\overline{\rm MS}$ mass at the scale $\mu$. 
If we use the heavy-quark pole mass $m$ instead of the $\overline{\rm MS}$ mass 
at the scale $\mu$, the order $\alpha_s$ correction in $T_H (x, \mu)$ involves a 
logarithm of $m_H^2/m^2$; instead, 
using the $\overline{\rm MS}$ mass in the Yukawa
coupling $y_Q$ ensures that $T_H$ depends only on the scale  $m_H$ and 
removes this logarithmic contribution from it~\cite{Koenig:2015pha}. 

The nonlocal operator ${\cal Q}^\alpha (x)$ has an anomalous dimension
known to NLO in $\alpha_s$ 
that allows us to resum logarithms of $m_H^2/m^2$ to NLL
accuracy~\cite{Lepage:1980fj, Mueller:1993hg, Mueller:1994cn, 
Vogelsang:1997ak, Hayashigaki:1997dn}. 
In order to resum the logarithms of  $m_H^2/m^2$ that appear in 
corrections of higher orders in $\alpha_s$ to the short-distance coefficients, 
we apply the factorization formula [Eq.~(\ref{eq:LCDAfac})] to the 
perturbative amplitudes $H \to Q \bar Q + \gamma$ and $H \to Q \bar Q g +
\gamma$. 
This involves calculating the decay constant and the LCDA with the meson state
$V$ replaced by the perturbative $Q \bar Q$ and $Q \bar Q g$ states. 
Then, each of the short-distance coefficients $c_n$ is given by 
a convolution of the hard part $T_H$ and a distribution in $x$ that satisfies
the same evolution equation as the nonlocal operator in 
Eq.~(\ref{eq:LCDAdef}). 
Equivalently, we can apply the NRQCD factorization formula to 
Eq.~(\ref{eq:LCDAdef}) to obtain an expression of the decay constant and the 
LCDA in terms of NRQCD LDMEs, so that~\cite{Jia:2008ep}  
\begin{eqnarray}
\label{eq:LCDANRQCDfac}%
f_V^\perp (\mu_0)
\phi_{V}^\perp (x,\mu_0)
&=& 
- \epsilon_\alpha (\lambda)
\langle V | 
{\cal Q}^\alpha (x)
| 0 \rangle
\nonumber \\
&=& 
\tilde c_0(x) 
\langle V | \psi^\dag \bm{\sigma} \cdot \bm{\epsilon} (\lambda) \chi | 0
\rangle 
+ \frac{\tilde c_{\bm{D}^2} (x)}{m^2} 
\langle V | \psi^\dag \bm{\sigma} \cdot \bm{\epsilon} (\lambda)
(-\tfrac{i}{2} \overleftrightarrow{\bm{D}})^2 \chi | 0 \rangle 
\nonumber\\ && 
+ \frac{\tilde c_{\bm{D}^4} (x)}{m^4} \langle V | \psi^\dag \bm{\sigma} \cdot
\bm{\epsilon} (\lambda)
(-\tfrac{i}{2} \overleftrightarrow{\bm{D}})^4 \chi | 0 \rangle 
\nonumber\\ && 
+ \frac{\tilde c_{\bm{D}^{(i} \bm{D}^{j)}} (x)}{m^2} 
\langle V | \psi^\dag 
\epsilon^i (\lambda) \sigma^j (-\tfrac{i}{2})^2 
\overleftrightarrow{\bm{D}}^{(i} \overleftrightarrow{\bm{D}}^{j)} 
\chi | 0 \rangle 
\nonumber\\ && 
+ \frac{\tilde c_B (x)}{m^2} \langle V | \psi^\dag g \bm{B} \cdot 
\bm{\epsilon} (\lambda)  \chi | 0 \rangle 
\nonumber\\ && 
+ \frac{\tilde c_{DE_0} (x)}{m^3} 
\langle V | \psi^\dag 
\bm{\sigma} 
\cdot 
\bm{\epsilon}
(\lambda)  
\tfrac{1}{3} 
( \overleftrightarrow{\bm{D}} \cdot g \bm{E} 
+g \bm{E} \cdot \overleftrightarrow{\bm{D}} ) 
\chi | 0 \rangle 
\nonumber\\ && 
+ \frac{\tilde c_{DE_1} (x)}{m^3} 
\langle V | \psi^\dag \bm{\epsilon} (\lambda)  \cdot 
\tfrac{1}{2} 
[\bm{\sigma} \times ( \overleftrightarrow{\bm{D}} \times g \bm{E}
- g \bm{E} \times \overleftrightarrow{\bm{D}} )] \chi | 0 \rangle , 
\end{eqnarray}
where the short-distance coefficients $\tilde c_n (x)$ 
are computed from the matching conditions 
that are similar to Eqs.~(\ref{eq:QQfactorization}) and 
(\ref{eq:QQgfactorization}), 
where the $c_n$ on the right-hand sides are replaced by 
$\tilde c_n (x)$, 
and the left-hand sides are replaced by 
$-\langle Q \bar Q(J^{PC}=1^{--}) | {\cal Q}^\alpha (x) | 0 \rangle$ 
and 
$-\langle Q \bar Q g(J^{PC}=1^{--})|{\cal Q}^\alpha (x)|0 \rangle$, 
respectively. 
In the matching conditions, 
the scale $\mu_0$ in Eq.~(\ref{eq:LCDANRQCDfac}) is the scale where the NRQCD
LDMEs on the right-hand side are defined. 
The factorization formula [Eq.~(\ref{eq:LCDAfac})] implies that 
\begin{equation}
\label{eq:fac_WC}%
c_n = - \frac{i}{2} e e_Q y_Q
\bm{\epsilon}^* (\lambda)
\cdot \bm{\epsilon}_\gamma^*
\int_0^1 dx\, T_H(x,\mu_0) \tilde c_n (x) 
+ O(m^2/m_H^2).
\end{equation}

It is worth noting that $\langle Q \bar Q | {\cal Q}^\alpha (x) | 0 \rangle$ 
and $\langle Q \bar Q g | {\cal Q}^\alpha (x) | 0 \rangle$
contain contributions from both negative and positive charge conjugation. 
From the fact that the charge conjugate of the operator 
${\cal Q}^\alpha (x)$ is given by $- {\cal Q}^\alpha (1-x)$, 
we can see that the contributions of negative charge conjugation 
to $\langle Q \bar Q | {\cal Q}^\alpha (x) | 0 \rangle$
and $\langle Q \bar Q g | {\cal Q}^\alpha (x) | 0 \rangle$
are given by the contribution symmetric in $x \leftrightarrow 1-x$. 
Since the hard part $T_H (x, \mu_0)$ is symmetric in $x \leftrightarrow 1-x$, the 
positive charge conjugation contribution of the LCDA, which is antisymmetric 
in $x \leftrightarrow 1-x$, does not contribute to the decay amplitude, 
consistently with the conservation of charge conjugation in the amplitude. 
Hence, in order to keep only contributions of negative charge 
conjugation in $\langle Q \bar Q | {\cal Q}^\alpha (x) | 0 \rangle$
and $\langle Q \bar Q g | {\cal Q}^\alpha (x) | 0 \rangle$, 
we just need to keep  contributions that are 
symmetric in $x \leftrightarrow 1-x$.

If we replace the meson state $V$ with the $Q \bar Q$ state, we obtain 
\begin{eqnarray}
\label{eq:LCDAQQ}%
- \langle Q \bar Q | {\cal Q}^\alpha (x) | 0 \rangle 
&=& 
- \int \frac{d \omega}{2 \pi} 
e^{-i (x-1/2) \omega \bar n \cdot P + i \omega \bar n \cdot q} 
\bar u (p_1) \bar n \!\!\!/ \gamma^\alpha_\perp v (p_2)
+O(g_s)
\nonumber \\
&=& 
- \frac{1}{\bar n \cdot P} \delta (x-1/2- \bar n \cdot q/\bar n \cdot P) 
\bar u (p_1) \bar n \!\!\!/ \gamma^\alpha_\perp v (p_2)+ O(g_s). 
\end{eqnarray}
We use the same strategy as the fixed-order calculation to 
obtain the contribution with $J^{PC}=1^{--}$; we express the 
Dirac bilinears in terms of the 3-momenta of the $Q$ and the $\bar Q$ in the $Q
\bar Q$ rest frame, and then we expand Eq.~(\ref{eq:LCDAQQ}) in powers of
$\bm{q}$ up to relative order $\bm{q^4}/m^4$. Then, Eq.~(\ref{eq:LCDAQQ}) is
given by a linear combination of the delta function $\delta(x-1/2)$ and its 
derivatives. In order to keep only the contribution with $C=-1$, we ignore the
odd derivatives of $\delta(x-1/2)$. 
The $J=1$ contribution is then obtained by reducing the 
Cartesian tensors of the form $\xi^\dag \eta q^i q^j \cdots q^k$ up to rank 4 and 
of the form $\xi^\dag \sigma^i \eta q^j \cdots q^k$ up to rank 5 using the
reduction method of Ref.~\cite{CoopeSnider}. 
We keep only the contribution with negative parity in order to obtain the
$J^{PC} = 1^{--}$ contribution. 
From the matching condition~(\ref{eq:LCDANRQCDfac}) we obtain 
\begin{subequations}
\begin{eqnarray}
{\tilde c}_0 (x) &=& \frac{1}{2 m} \delta(x-1/2) , 
\\
{\tilde c}_{\bm{D}^2} (x) &=& 
\frac{1}{m}
\left[ -\frac{5}{12} \delta(x-1/2) + \frac{1}{48} \delta^{(2)} (x-1/2) \right],
\\
{\tilde c}_{\bm{D}^{(i} \bm{D}^{j)}} (x) &=& 
\frac{1}{m}
\left[ \frac{1}{4} \delta(x-1/2) - \frac{1}{80} \delta^{(2)} (x-1/2) \right],
\\
{\tilde c}_{\bm{D}^4} (x) &=& 
\frac{1}{m} 
\left[ \frac{19}{48} \delta(x-1/2) - \frac{19}{480} \delta^{(2)} (x-1/2) 
+ \frac{1}{3840} \delta^{(4)} (x-1/2) \right],
\\
{\tilde c}_{\bm{D}^2 \bm{D}^{(i} \bm{D}^{j)}} (x) &=& 
\frac{1}{m} 
\left[ -\frac{5}{16} \delta(x-1/2) + \frac{1}{32} \delta^{(2)} (x-1/2) 
- \frac{1}{4480} \delta^{(4)} (x-1/2) \right] . 
\end{eqnarray}
\end{subequations}
The short-distance coefficients $\tilde c_0 (x)$ and 
$\tilde c_{\bm{D}^2} (x)$ agree with known results in 
Ref.~\cite{Bodwin:2014bpa}.
The agreement with the fixed-order calculation can be easily verified 
using Eq.~(\ref{eq:fac_WC}). 

To compute the remaining short-distance coefficients 
corresponding to the color-octet LDMEs, 
we replace the meson state $V$ 
with the $Q \bar Q g$ state to obtain 
\begin{eqnarray}
\label{eq:LCDAQQg}%
\hspace{-8ex}
- \langle Q \bar Q g | {\cal Q}^\alpha (x) | 0 \rangle 
&=& 
- \int \frac{d \omega}{2 \pi} 
e^{-i (x-1/2) \omega \bar n \cdot P + i \omega \bar n \cdot (p_1-p_2)/2} 
\nonumber \\ && \hspace{3ex} \times 
\bigg[ 
e^{i \omega \bar n \cdot k_g/2} 
\bar u(p_1) (-i g_s \epsilon\!\!\!/_g^* T^a) 
\frac{i (p\!\!\!/_1+k\!\!\!/_g+m)}{(p_1+k_g)^2-m^2+i \varepsilon}
\bar n \!\!\!/ \gamma^\alpha_\perp v(p_2) 
\nonumber \\ && \hspace{5ex} 
+ 
e^{-i \omega \bar n \cdot k_g/2} 
\bar u(p_1) \bar n \!\!\!/ \gamma^\alpha_\perp 
\frac{i (-p\!\!\!/_2-k\!\!\!/_g+m)}{(p_2+k_g)^2-m^2+i \varepsilon}
(-i g_s \epsilon\!\!\!/_g^* T^a) 
v(p_2) 
\nonumber \\ && \hspace{5ex} 
+ 
e^{i \omega \bar n \cdot k_g/2} 
(i g_s) \frac{i \bar n \cdot \epsilon_g^*}{\bar n \cdot k_g}
\bar u(p_1) 
\bar n\!\!\!/ \gamma^\alpha_\perp  T^a v(p_2)
\nonumber \\ && \hspace{5ex} 
+ 
e^{-i \omega \bar n \cdot k_g/2} 
(-i g_s)
\frac{i \bar n \cdot \epsilon_g^*}{\bar n \cdot k_g}
\bar u(p_1) \bar n\!\!\!/ \gamma^\alpha_\perp T^a v(p_2)
\bigg] + O(g_s^2)
\nonumber \\ &=& 
\frac{i g_s}{\bar n \cdot P} 
\delta\left(x-1/2-\tfrac{\bar n \cdot (p_1-p_2+k_g)}{2 \bar n \cdot P}\right)
\bar u(p_1) \epsilon\!\!\!/_g^* T^a 
\frac{i (p\!\!\!/_1+k\!\!\!/_g+m)}{2p_1\cdot k_g+i \varepsilon}
\bar n \!\!\!/ \gamma^\alpha_\perp v(p_2) 
\nonumber \\ && 
+ 
\frac{i g_s}{\bar n \cdot P} 
\delta\left(x-1/2- \tfrac{\bar n \cdot (p_1-p_2-k_g)}{2 \bar n \cdot P}\right)
\bar u(p_1) \bar n \!\!\!/ \gamma^\alpha_\perp 
\frac{i (-p\!\!\!/_2-k\!\!\!/_g+m)}{2p_2\cdot k_g+i \varepsilon}
\epsilon\!\!\!/_g^* T^a
v(p_2)
\nonumber \\ && 
- \frac{i g_s}{\bar n \cdot P} 
\delta\left(x-1/2- \tfrac{\bar n \cdot (p_1-p_2+k_g)}{2 \bar n \cdot P}\right) 
\frac{i \bar n \cdot \epsilon_g^*}{\bar n \cdot k_g}
\bar u(p_1) 
\bar n\!\!\!/ \gamma^\alpha_\perp T^a v(p_2)
\nonumber \\ && 
- \frac{i g_s}{\bar n \cdot P}
\delta\left(x-1/2- \tfrac{\bar n \cdot (p_1-p_2-k_g)}{2 \bar n \cdot P}\right) 
\frac{-i \bar n \cdot \epsilon_g^*}{\bar n \cdot k_g}
\bar u(p_1) \bar n\!\!\!/ \gamma^\alpha_\perp T^a v(p_2) + O(g_s^2). 
\end{eqnarray}
After expressing the amplitude in terms of the 3-momenta of the
$Q$, $\bar Q$ and $g$, we expand the amplitude in powers of $\bm{q}_1$ and
$\bm{q}_2$ up to relative order $\bm{q}_1^2 / m^2$, $\bm{q}_2^2 / m^2$,
and $|\bm{q}_1| |\bm{q}_2| / m^2$. 
The resulting expression for the amplitude is then a linear combination of the
Cartesian tensors built from $\xi^\dag \bm{\sigma} \eta$,
$\bm{\epsilon}_g^*$, $\bm{q}_1$, and $\bm{q}_2$ up to rank 4.
We obtain the matching condition for $Q \bar Q g$ with $J^{PC} = 1^{--}$ 
by keeping only the contributions symmetric in 
$x \leftrightarrow 1-x$, i.e., $C=-1$,
reducing the Cartesian tensors to $J=1$, and keeping only the negative parity
contributions. 
The resulting matching condition leads to the following 
short-distance coefficients 
\begin{subequations}
\begin{eqnarray}
{\tilde c}_{B} (x) &=& 
\frac{1}{2 m} \delta(x-1/2), 
\\
{\tilde c}_{DE_0} (x) &=& 
\frac{1}{m} \left[ - \frac{7}{16} \delta (x-1/2) 
+ \frac{1}{128} \delta^{(2)} (x-1/2) \right] , 
\\
{\tilde c}_{DE_1} (x) &=& 
\frac{1}{m} \left[ - \frac{1}{8} \delta(x-1/2) 
- \frac{1}{128}  \delta^{(2)} (x-1/2) \right] , 
\\
{\tilde c}_{DE_1'} (x) &=& 
\frac{1}{m} \left[ - \frac{3}{8} \delta(x-1/2) 
- \frac{1}{32} \delta^{(2)} (x-1/2) \right] ,
\\
{\tilde c}_{DE_2} (x) &=& 
\frac{1}{m} \left[ \frac{1}{8} \delta(x-1/2) - 
\frac{1}{640} \delta^{(2)} (x-1/2) \right]. 
\end{eqnarray}
\end{subequations}
The agreement with the fixed-order calculation can be easily verified 
using Eq.~(\ref{eq:fac_WC}). 

By integrating the short-distance coefficients $\tilde{c}_n (x)$ over
$x$, we
obtain an expression for $f_V^\perp$ valid up to relative order $v^4$. 
If we include the correction of relative order $\alpha_s v^0$ 
in the $\overline{\rm MS}$ scheme 
from Ref.~\cite{Wang:2013ywc}, we obtain 
\begin{eqnarray}
\label{eq:fVperpresult}%
f_V^\perp (\mu_0)
&=& 
\frac{\langle V | \psi^\dag \bm{\sigma} \cdot \bm{\epsilon} (\lambda) \chi 
| 0 \rangle}{2 m} \bigg[
1 - \frac{\alpha_s (\mu_0) C_F}{4 \pi} 
\left( \log \frac{\mu^2_0}{m^2} + 8 \right) 
-\frac{5}{6} \langle v^2_S \rangle_V 
+ \frac{19}{24} \langle v^4_S \rangle_V
\nonumber\\ && \hspace{5ex}
- \frac{7}{8} \langle DE_0 \rangle_V
+ \frac{1}{2} \langle v^2_D \rangle_V 
+ \langle B \rangle_V
- \frac{1}{4} \langle DE_1 \rangle_V
+O(v^5)
\bigg]. 
\end{eqnarray}
The corrections at relative order $v^2$ agree with Ref.~\cite{Bodwin:2014bpa}.
The logarithm of $\mu_0^2/m^2$ in Eq.~(\ref{eq:fVperpresult}) is the remnant of
the renormalization of the decay constant in the $\overline{\rm MS}$ scheme. 

Since the short-distance coefficients $\tilde{c}_n (x)$ 
at leading order in $\alpha_s$ 
are linear combinations of $\delta(x-1/2)$, $\delta^{(2)} (x-1/2)$ and
$\delta^{(4)} (x-1/2)$, the LCDA can be written as 
\begin{equation}
\label{eq:LCDAresult1}%
\phi_V^\perp (x,\mu_0) = 
\phi_V^\perp{}^{(0)} (x,\mu_0) + 
\phi_V^\perp{}^{(\alpha_s)} (x, \mu_0) + 
\phi_V^\perp{}^{(2)} (x, \mu_0) 
+ \phi_V^\perp{}^{(4)} (x, \mu_0), 
\end{equation}
where 
\begin{subequations}
\label{eq:LCDAresult2}%
\begin{eqnarray}
\phi_V^\perp{}^{(0)} (x,\mu_0) &=& \delta(x-1/2), \\
\label{eq:LCDAresult2b}%
\phi_V^\perp{}^{(2)} (x,\mu_0) &=&
\bigg[ \frac{1}{3} \langle v^2_S \rangle_V
+\frac{5}{18} \left(\langle v^2_S \rangle_V\right)^2 
- \frac{19}{30} \langle v^4_S \rangle_V 
\nonumber \\ && 
+ \frac{1}{8} \langle DE_0 \rangle 
- \frac{1}{5} \langle v^2_D \rangle_V 
- \frac{1}{8} \langle DE_1 \rangle_V \bigg] 
\frac{\delta^{(2)} (x-1/2)}{8},
\\ 
\phi_V^\perp{}^{(4)} (x,\mu_0) &=& 
\frac{1}{5} \langle v^4_S \rangle_V \frac{\delta^{(4)} (x-1/2)}{384}, 
\end{eqnarray}
\end{subequations}
and $\phi_V^\perp{}^{(\alpha_s)} (x,\mu_0)$ 
is the order $\alpha_s v^0$ correction in the $\overline{\rm MS}$ scheme 
given by~\cite{Wang:2013ywc} 
\begin{eqnarray}
\label{eq:LCDAoneloop}%
\phi_V^\perp{}^{(\alpha_s)} (x,\mu_0) &=& 
\frac{\alpha_s (\mu_0) C_F}{4 \pi} 
\theta(1-2 x) 
\bigg\{ \left[ 
\frac{8 x}{1-2 x} \left( \log \frac{\mu_0^2}{m^2 (1-2 x)^2}-1 \right) \right]_+
\\ \nonumber && \hspace{22ex} 
+ \left[ \frac{16 x (1-x)}{(1-2 x)^2} \right]_{++} \bigg\}
+ (x \leftrightarrow 1-x),
\end{eqnarray}
where the plus and plus-plus distributions are defined by 
\begin{subequations}
\begin{eqnarray}
&& \int_0^1 dx\, f(x) [g(x)]_+ = \int_0^1 dx \, [f(x)-f(1/2)] g(x), 
\\
&& \int_0^1 dx\, f(x) [g(x)]_{++} = 
\int_0^1 dx \, [f(x)-f(1/2)-f'(1/2) (x-1/2)] g(x). 
\end{eqnarray}
\end{subequations}
The logarithm of $\mu_0^2/m^2$ in Eq.~(\ref{eq:LCDAoneloop}) is the remnant of
the renormalization of the LCDA in the $\overline{\rm MS}$ scheme. 
Note that 
\begin{equation}
\int_0^1 \frac{dx}{x (1-x)}  \delta (x-1/2) 
= \int_0^1 \frac{dx}{x (1-x)}
\frac{\delta^{(2)} (x-1/2)}{8} 
= \int_0^1 \frac{dx}{x (1-x)}
\frac{\delta^{(4)} (x-1/2)}{384} = 4. 
\end{equation}

The decay constant and the LCDA computed in 
Eqs.~(\ref{eq:fVperpresult}, \ref{eq:LCDAresult1}) allow us to resum
logarithms of $m_H^2/m^2$ that appear in the QCD corrections to the direct
amplitude to all orders in $\alpha_s$. 
This is accomplished by solving the evolution equation 
\begin{equation}
\mu^2 \frac{\partial}{\partial \mu^2} 
f_V^\perp (\mu) \phi_V^\perp (x, \mu)
= f_V^\perp (\mu) \frac{\alpha_s (\mu) C_F}{2 \pi} 
\int_0^1 dy \, V_T (x,y)  \phi_V^\perp (y, \mu), 
\end{equation}
where the evolution kernel $V_T(x,y)$ is currently known to 
NLO accuracy in $\alpha_s$~\cite{Lepage:1980fj, Mueller:1993hg, Mueller:1994cn, 
Vogelsang:1997ak, Hayashigaki:1997dn}. 
We obtain $f_V^\perp (\mu) \phi_V^\perp (x, \mu)$ at scale $\mu$ 
from $f_V^\perp (\mu_0) \phi_V^\perp (x, \mu_0)$ at scale $\mu_0$ by solving
this evolution equation. 
The direct amplitude with logarithms of $m_H^2/m^2$ resummed to all orders in 
$\alpha_s$ is then given by Eq.~(\ref{eq:LCDAfac}), by setting 
$\mu \sim m_H$ and $\mu_0 \sim m$, so that $T_H (x,\mu)$ is free of logarithms
of $m_H^2/m^2$.\footnote{
Equivalently, one may resum logarithms of 
$m_H^2/m^2$ in the hard part $T_H(x,\mu)$, which may be conceptually closer to
the effective field theory logic.  
}
The accuracy of the resummation is limited by the accuracy of
the evolution kernel; since the evolution kernel is known up to NLO accuracy,
we can resum the logarithms to NLL accuracy.
The resummation can be carried out using the Gegenbauer polynomials 
$C_n^{(3/2)} (2 x-1)$, which are the eigenfunctions of the LO evolution
kernel~\cite{Brodsky:1980ny}. 
The convolution in Eq.~(\ref{eq:LCDAfac}) is given by 
\begin{equation}
f_V^\perp (\mu) \int_0^1 dx\, T_H (x,\mu) \phi_V^\perp (x, \mu) = 
f_V^\perp (\mu) \sum_{n=0}^\infty \hat T_H (n,\mu) \hat \phi_V^\perp (n,\mu), 
\end{equation}
where $\hat T_H (n,\mu)$ and $\hat \phi_V^\perp (n,\mu)$ are Gegenbauer 
moments defined by 
\begin{subequations}
\begin{eqnarray}
\hat T_H (n,\mu) &=& \int_0^1 dx\, x (1-x) C_n^{(3/2)} (2 x-1) T_H (x,\mu), \\
\hat \phi_V^\perp (n,\mu) &=& 
\frac{4 (2 n+3)}{(n+1) (n+2)} 
\int_0^1 dx \, C_n^{(3/2)} (2 x-1) \phi_V^\perp (x,\mu). 
\end{eqnarray}
\end{subequations}
The solution of the evolution equation in terms of Gegenbauer moments 
leads to~\cite{Mueller:1993hg, Mueller:1994cn, 
Vogelsang:1997ak, Hayashigaki:1997dn}
\begin{subequations}
\label{eq:conv_sum}%
\begin{eqnarray}
\label{eq:conv_sum_a}%
\hat \phi_V^\perp (n_1,\mu) 
&=& \sum_{n_2=0} ^\infty U_{n_1 n_2} (\mu, \mu_0) 
\hat \phi_V^\perp (n_2, \mu_0), 
\\
f_V^\perp (\mu) &=& U_{f_V^\perp} (\mu, \mu_0) f_V^\perp (\mu_0). 
\end{eqnarray}
\end{subequations}
Explicit expressions of $ U_{f_V^\perp} (\mu, \mu_0)$ and 
$U_{n_1 n_2} (\mu, \mu_0)$ can be found in Ref.~\cite{Bodwin:2016edd}. 
We note that $U_{f_V} (\mu, \mu_0)$ and
$U_{n_1 n_2} (\mu, \mu_0)$ depend only on $\mu$, $\mu_0$ and the evolution 
kernel, and are independent of $f_V^\perp (\mu_0)$ or $\phi_V^\perp (x,\mu_0)$. 
When $\phi_V^\perp (x,\mu_0)$
contains singular distributions such as the delta function and their 
derivatives, the sum in Eq.~(\ref{eq:conv_sum_a}) can converge badly. 
The authors of Refs.~\cite{Bodwin:2016edd, Bodwin:2017wdu} developed a method
to overcome the problem of nonconvergence by defining the sum over $n_2$ in 
Eq.~(\ref{eq:conv_sum_a}) as an Abel sum, so that 
\begin{eqnarray}
&& 
f_V^\perp (\mu) \int_0^1 dx\, T_H (x,\mu) \phi_V^\perp (x, \mu)
\nonumber \\ 
&& \hspace{5ex} =  
\lim_{z\to 1}
U_{f_V^\perp} (\mu, \mu_0) 
f_V^\perp (\mu_0)
\sum_{n_1=0}^\infty 
\sum_{n_2=0} ^\infty \hat T_H (n_1,\mu) 
U_{n_1 n_2} (\mu, \mu_0) \hat \phi_V^\perp (n_2, \mu_0) z^{n_2}. 
\end{eqnarray}
The numerical value of the Abel sum is
then evaluated by using Pad\'e approximants of the truncated series. 
We employ this Abel-Pad\'e method~\cite{Bodwin:2016edd, Bodwin:2017wdu} 
to compute the convolution in 
Eq.~(\ref{eq:conv_sum_a}). 

It is more convenient to express the direct amplitude with resummed
logarithms in terms of the
decay constant and the LCDA convolved with the hard part, i.e.
Eq.~(\ref{eq:LCDAfac}), than to 
resum the logarithms for each of the short-distance
coefficients in Eqs.~(\ref{eq:WC_singlet}) and (\ref{eq:WC_octet}). 
We note that the resummation of logarithms can be carried out separately for 
the decay constant and the individual contributions to the LCDA in 
Eq.~(\ref{eq:LCDAresult1}), so that 
\begin{eqnarray}
\label{eq:amplitude_direct_resummed}%
i {\cal M}_{\rm dir} (H \to V + \gamma) &=& 
\frac{i}{2} e e_Q y_Q
\epsilon_\gamma^*{} \cdot \epsilon^* (\lambda)
\sum_{n = 0,2,4,\alpha_s}
U_{f_V^\perp} (\mu, \mu_0)
f_V^\perp (\mu_0) 
\int_0^1 dx\, T_H (x,\mu) 
\phi_V^\perp{}^{(n)} (x,\mu)
\nonumber \\ && + O(m^2/m_H^2), 
\end{eqnarray}
where 
\begin{equation}
\int_0^1 dx\, T_H (x,\mu) \phi_V^\perp{}^{(n)} (x, \mu)
= 
\sum_{n_1=0}^\infty 
\sum_{n_2=0} ^\infty \hat T_H (n_1,\mu) 
U_{n_1 n_2} (\mu, \mu_0) \hat \phi_V^\perp{}^{(n)} (n_2, \mu_0). 
\end{equation}

\section{Calculation of the indirect amplitude}
\label{sec:indirect}%

The indirect amplitude proceeds from $H \to \gamma^* \gamma$, followed by
$\gamma^* \to V$. Since we work in the limit $m_V^2 / m_H^2 \to 0$, 
the partial amplitude $H \to \gamma^* \gamma$ can be replaced by the 
decay amplitude of the Higgs boson to two photons. 
Then, the indirect amplitude is given by~\cite{Bodwin:2013gca} 
\begin{equation}
i {\cal M}_{\rm ind} (H \to V + \gamma)
= - i \epsilon_\gamma^* \cdot \epsilon^* (\lambda)
\frac{e_Q f_V^\parallel 
\sqrt{4 \pi \alpha(\mu_0)}}{m_V} 
\left[ 
16 \pi m_H 
\frac{\alpha (\mu_0)}{\alpha(0)}
\Gamma(H \to \gamma \gamma)
\right]^{\frac{1}{2}}, 
\end{equation}
where the factor $16 \pi m_H$ 
compensates for the factors that 
are necessary to obtain the decay rate $\Gamma(H \to \gamma \gamma)$ 
from the squared amplitude.
We take the scale of the QED coupling at the vertices associated with the 
virtual photon with virtuality $m_V$ to be $\mu_0$, and the scale of the vertex
for the real photon to be $0$. 
The factor $\alpha (\mu_0)/\alpha(0)$ replaces one QED coupling
constant at scale 0 in the $H \to \gamma \gamma$ amplitude with 
the QED coupling constant at scale $\mu_0$. 
We ignore the small imaginary part in the $H \to \gamma \gamma$
amplitude~\cite{Koenig:2015pha, Bodwin:2016edd}. 
The decay constant $f_V^{\parallel}$ is defined by 
\begin{equation}
f_V^{\parallel} =
- \frac{1}{m_V} \langle V | \bar Q \epsilon\!\!\!/ (\lambda) Q | 0 \rangle. 
\end{equation}
Since $f_V^{\parallel}$ is a conserved current in QCD, it 
does not undergo renormalization.
Hence, the indirect amplitude is free of logarithms of
$m_H^2/m^2$ in the limit $m_V^2 / m_H^2 \to 0$ if one ignores the
higher-order electroweak corrections to the indirect amplitude. 
We can express $f_V^\parallel$ in terms of NRQCD LDMEs up to relative order
$v^4$
using the same techniques we employed to compute the direct amplitude and the
decay constant $f_V^\perp$. We obtain 
\begin{eqnarray}
\label{eq:fvpar}%
f_V^{\parallel}
&=& 
\frac{\langle V | \psi^\dag \bm{\sigma} \cdot \bm{\epsilon} (\lambda) 
\chi | 0 \rangle 
}{2 m} \bigg(
1 - 8 \frac{\alpha_s (\mu_0) C_F}{4 \pi} 
-\frac{2}{3} \langle v^2_S \rangle_V
+ \frac{7}{12} \langle v^4_S \rangle_V 
\nonumber\\ && \hspace{5ex}
- \frac{5}{8} \langle DE_0 \rangle_V 
- \frac{1}{2} \langle v^2_D \rangle_V
+ \frac{1}{2} \langle B \rangle_V 
- \frac{1}{4} \langle DE_1 \rangle_V 
+O(v^5) 
\bigg), 
\end{eqnarray}
where we included the order $\alpha_s v^0$ correction computed in
Refs.~\cite{Barbieri:1975ki, Celmaster:1978yz}. 
We note that $f_V^\parallel$ is positive at leading order in $\alpha_s$ and 
$v$. If we assume $f_V^\parallel>0$, an accurate numerical value for 
$f_V^\parallel$ can be obtained from the leptonic decay rate 
\begin{equation}
\Gamma (V \to \ell^+ \ell^-) = \frac{8 \pi}{3} \alpha^2 (\mu_0) e_Q^2 
\left|f_V^{\parallel} \right|^2,
\end{equation}
so that 
\begin{equation}
\label{eq:amplitude_indirect}%
i {\cal M}_{\rm ind} (H \to V + \gamma)
= - i \epsilon_\gamma^* \cdot \epsilon^* (\lambda)
\frac{e_Q}{|e_Q|} 
\frac{\sqrt{24 \pi m_H}}{m_V} 
\left[ 
\frac{ 
\Gamma(V \to \ell^+ \ell^-)
\Gamma(H \to \gamma \gamma)
}{\alpha (0)} 
\right]^{\frac{1}{2}}. 
\end{equation}
Note that $i {\cal M}_{\rm ind}$ and $i {\cal M}_{\rm dir}$ have opposite 
signs, so that when calculating the decay rate, the direct and indirect
amplitudes interfere destructively~\cite{Bodwin:2013gca}.

\section{Numerical Results}
\label{sec:results}%

We now present our numerical results for the Higgs decay rate into $V +
\gamma$ for $V= J/\psi$ and $\Upsilon(nS)$ for $n=1,2$, and 3 
based on our calculation of the direct amplitude to relative order
$v^4$ accuracy. 
As we have mentioned in Sec.~\ref{sec:introduction}, we do not consider the
$\psi(2S)$ state, because there are no available estimates of the 
relevant NRQCD matrix elements that account for open-flavor threshold effects 
and nonrelativistic corrections in a complete and model-independent way. 

Our expressions for the direct and indirect amplitudes are given in 
Eqs.~(\ref{eq:amplitude_direct_resummed}) and~(\ref{eq:amplitude_indirect}), 
respectively. 
If we write ${\cal M}_{\rm dir} = \epsilon_\gamma^* \cdot \epsilon^* (\lambda) 
{\cal A}_{\rm dir} / \sqrt{\Phi}$ and 
${\cal M}_{\rm ind} = \epsilon_\gamma^* \cdot \epsilon^* (\lambda) 
{\cal A}_{\rm ind} / \sqrt{\Phi}$, the decay rate 
$\Gamma(H \to V + \gamma)$ is given by 
\begin{equation}
\label{eq:decayrate}%
\Gamma(H \to V + \gamma)
= 
| {\cal A}_{\rm dir} + {\cal A}_{\rm ind} | ^2, 
\end{equation}
where $\Phi$ is the phase-space and normalization factor given by 
\begin{equation}
\label{eq:phase_space}%
\Phi = \frac{1}{2 m_H} \frac{m_V (m_H^2 - m_V^2)}{2 \pi m_H^2} . 
\end{equation}
We present our numerical results for ${\cal A}_{\rm dir}$ and 
${\cal A}_{\rm ind}$ in the following sections. 
Then, using the resulting values of 
${\cal A}_{\rm dir}$ and ${\cal A}_{\rm ind}$, we compute the decay rate 
$\Gamma(H \to V + \gamma)$ from the formula in Eq.~(\ref{eq:decayrate}).

\subsection{Indirect amplitude} 

We compute the numerical values for the indirect amplitude 
using Eq.~(\ref{eq:amplitude_indirect}). 
We compute the decay constant $f_V^\parallel$ from the measured leptonic decay
rate: 
\begin{equation}
\label{eq:fvpar_from_rate}%
f_V^{\parallel} =
\bigg[ \frac{3 \, \Gamma(V \to \ell^+ \ell^-) }{8 \pi \alpha^2 (\mu_0) e_Q^2} 
\bigg]^{\frac{1}{2}}, 
\end{equation}
where we take $\mu_0 = m_V$. Then 
\begin{equation}
\label{eq:amp_indirect}%
{\cal A}_{\rm ind} = 
- 
\sqrt{\Phi}
\frac{e_Q}{|e_Q|} 
\frac{\sqrt{24 \pi m_H}}{m_V} 
\left[ 
\frac{ 
\Gamma(V \to \ell^+ \ell^-)
\Gamma(H \to \gamma \gamma)
}{\alpha (0)} 
\right]^{\frac{1}{2}}. 
\end{equation}
We note that the resulting expression for 
${\cal A}_{\rm ind}$ does not depend on $\alpha(\mu_0)$. 

We use the following input parameters to compute ${\cal A}_{\rm ind}$ 
numerically. 
We take the PDG value for the Higgs mass $m_H = 125.18 \pm
0.16$~\cite{Tanabashi:2018oca}, 
and the numerical value for the Higgs two-photon decay rate to be 
$\Gamma(H \to \gamma \gamma) = 9.34 \times 10^{-6}$~GeV, which is computed
from the tabulated results of the two-photon branching ratio and the total
decay rate of the Higgs boson in Refs.~\cite{Dittmaier:2011ti, Dittmaier:2012vm}. 
We also take the PDG values for the meson masses $m_V$ and use the measured 
partial widths $\Gamma(V \to e^+ e^-)$ for the leptonic decay
rates~\cite{Tanabashi:2018oca}. 
We list the values for $m_V$ and $\Gamma(V \to \ell^+ \ell^-)$ that we use to
compute the indirect amplitude in Table.~\ref{table:inputs}. 
The QED coupling constant at scale 0 is taken to be 
$\alpha(0) = 1/137.036$. 

\begin{table}[t]
\begin{center}
\begin{ruledtabular}
\begin{tabular}{ccc}
$V$ &
$m_V$~(GeV) & 
$\Gamma(V \to \ell^+ \ell^-)$~(keV) 
\\
\hline
$J/\psi$ & $3.0969$ & 
$5.55 \pm 0.14 \pm 0.02$
\\
$\Upsilon(1S)$ & $9.4603$ & 
$1.340 \pm 0.018$
\\
$\Upsilon(2S)$ & $10.02326$ & 
$0.612 \pm 0.011$
\\
$\Upsilon(3S)$ & $10.3352$ & 
$0.443 \pm 0.008$
\\
\end{tabular}
\end{ruledtabular}
\caption{\label{table:inputs}%
Values for the meson masses $m_V$ and the leptonic decay rates 
$\Gamma(V \to \ell^+ \ell^-)$ used in the numerical calculation of the direct and
indirect amplitudes. 
All values are taken from Ref.~\cite{Tanabashi:2018oca}.  
}
\end{center}
\end{table}

We consider the following sources of uncertainties in ${\cal A}_{\rm ind}$. 
The uncertainty in the decay rate $\Gamma(H \to \gamma \gamma)$ is taken to be
$0.01$ times the central value, as estimated in Ref.~\cite{Dittmaier:2011ti}
from the higher-order corrections to the decay rate. 
We consider the experimental uncertainties in
$\Gamma(V \to \ell^+ \ell^-)$. We estimate the uncalculated correction of
relative order $m_V^2/m_H^2$ to be $m_V^2/m_H^2$ of the central value. 
We ignore the negligibly small uncertainties 
in $m_H$ and $m_V$ compared to other sources of uncertainties. 
We add the uncertainties in quadrature.
Our numerical results for ${\cal A}_{\rm ind}$ are shown in
Table.~\ref{table:amplitudes}. 
We note that the uncertainties in ${\cal A}_{\rm ind}$ are less than 2\% of the 
central values. 

We can compare our results for ${\cal A}_{\rm ind}$ 
with a previous calculation in Ref.~\cite{Bodwin:2017wdu}. 
Our calculation is equivalent to the one in Ref.~\cite{Bodwin:2017wdu}, except that
we use an updated value of the measured Higgs mass from 
Ref.~\cite{Tanabashi:2018oca}, which has smaller
uncertainties than what was employed in Ref.~\cite{Bodwin:2017wdu}.
Our results for ${\cal A}_{\rm ind}$ in Table.~\ref{table:amplitudes}
are compatible with those in Ref.~\cite{Bodwin:2017wdu} within uncertainties.

\subsection{Direct amplitude} 

We compute the numerical values of the direct amplitude 
using Eq.~(\ref{eq:amplitude_direct_resummed}), so that 
\begin{eqnarray}
\label{eq:ampdir}%
{\cal A}_{\rm dir} &=& 
\frac{1}{2} 
\sqrt{\Phi}
e e_Q y_Q
\sum_{n = 0,2,4,\alpha_s}
U_{f_V^\perp} (\mu, \mu_0) f_V^\perp (\mu_0) 
\int_0^1 dx \, T_H (x,\mu) 
\phi_V^\perp{}^{(n)} (x,\mu).
\end{eqnarray}
We now discuss our strategy to compute Eq.~(\ref{eq:ampdir}) numerically. 
The decay constant $f_V^\perp (\mu_0)$ depends on the 
LDME $\langle V | \psi^\dag \bm{\sigma} \cdot \bm{\epsilon}
(\lambda) \chi | 0 \rangle$ and the ratios of LDMEs 
$\langle v_S^2 \rangle_V$, 
$\langle v_S^4 \rangle_V$, 
$\langle v_D^2 \rangle_V$, 
$\langle DE_0 \rangle_V$, 
$\langle B \rangle_V$, and 
$\langle DE_1 \rangle_V$, see Eq.~(\ref{eq:fVperpresult}).
The dependence on the leading-order LDME 
$\langle V | \psi^\dag \bm{\sigma} \cdot \bm{\epsilon}
(\lambda) \chi | 0 \rangle$ 
can be eliminated by rewriting the decay constant as 
\begin{eqnarray}
\label{eq:fVratio}%
f_V^\perp (\mu_0) = 
\frac{f_V^\perp (\mu_0)}{f_V^{\parallel}} f_V^{\parallel} 
&=& 
f_V^{\parallel} \bigg[ 
1 - \frac{\alpha_s (\mu_0) C_F}{4 \pi} 
\log \frac{\mu_0^2}{m^2} 
-\frac{1}{6} \langle v^2_S \rangle_V
-\frac{1}{9} \left(\langle v^2_S \rangle_V \right)^2
+ \frac{5}{24} \langle v^4_S \rangle_V
\nonumber\\ && \hspace{5ex}
- \frac{1}{4} \langle DE_0 \rangle_V
+ \langle v^2_D \rangle_V 
+ \frac{1}{2} \langle B \rangle_V +O(v^5) \bigg] , 
\end{eqnarray}
and by obtaining the numerical value for $f_V^{\parallel}$ 
from the measured leptonic decay rate using Eq.~(\ref{eq:fvpar_from_rate}). 
The $\phi_V^\perp{}^{(2)} (x,\mu)$ term in the LCDA
depends on the ratios $\langle v_S^2 \rangle_V$,
$\langle v_S^4 \rangle_V$, $\langle v_D^2 \rangle_V$, $\langle DE_0 \rangle_V$,
and $\langle DE_1 \rangle_V$, see Eq.~(\ref{eq:LCDAresult2b}), 
while the remaining contributions 
in Eq.~(\ref{eq:LCDAresult1}) depend only on
$\langle v_S^4 \rangle_V$. 
We can eliminate the ratios $\langle B \rangle_V$ and $\langle DE_0 \rangle_V$ 
in $f_V^\perp (\mu_0)$ and $\phi_V^\perp{}^{(2)} (x,\mu)$ 
by using the Gremm-Kapustin relations in Eqs.~(\ref{eq:Gremm}). 
We obtain 
\begin{eqnarray}
\label{eq:fVratio_GK}%
f_V^\perp (\mu_0)  
&=& 
f_V^{\parallel} \bigg[ 
1 - \frac{\alpha_s (\mu_0) C_F}{4 \pi} 
\log \frac{\mu_0^2}{m^2} 
- \frac{m_V - 2 m}{2 m} 
+ \left( \frac{1}{3} + \frac{m_V-2 m}{6 m} \right) \langle v^2_S \rangle_V
\nonumber\\ && \hspace{5ex}
-\frac{1}{9} \left(\langle v^2_S \rangle_V \right)^2
- \frac{1}{12} \langle v^4_S \rangle_V
+ \langle v^2_D \rangle_V 
+ \frac{1}{8} \langle DE_1 \rangle_V
+O(v^5) \bigg] , 
\end{eqnarray}
and 
\begin{eqnarray}
\label{eq:LCDAresultGK}%
\phi_V^\perp{}^{(2)} (x,\mu_0) &=&
\bigg[ \left(\frac{1}{3}- \frac{m_V-2 m}{12 m} \right) \langle v^2_S \rangle_V
+\frac{5}{18} \left(\langle v^2_S \rangle_V\right)^2 
- \frac{11}{20} \langle v^4_S \rangle_V 
\nonumber \\ && \hspace{3ex}
- \frac{1}{5} \langle v^2_D \rangle_V 
- \frac{1}{8} \langle DE_1 \rangle_V + O(v^5) \bigg] 
\frac{\delta^{(2)} (x-1/2)}{8}.
\end{eqnarray}
When computing the convolution in Eq.~(\ref{eq:ampdir}), the terms
$f_V^\perp (\mu_0) \phi_V^\perp{}^{(n)} (x, \mu_0)$ 
for $n=\alpha_s$, 2, and 4 can contain cross terms that go beyond our current 
level of accuracy. 
In order to avoid such contributions, 
we ignore the cross terms that contribute to the direct amplitude beyond 
relative order $\alpha_s v^0$ and $v^4$, so that 
\begin{subequations}
\label{eq:LCDAresult3}%
\begin{eqnarray}
f_V^\perp (\mu_0)  
\phi_V^\perp{}^{(0)} (x,\mu_0) &=&
f_V^{\parallel} \bigg[ 
1 - \frac{\alpha_s (\mu_0) C_F}{4 \pi} 
\log \frac{\mu_0^2}{m^2} 
- \frac{m_V - 2 m}{2 m} 
+ \left( \frac{1}{3} + \frac{m_V-2 m}{6 m} \right) \langle v^2_S \rangle_V
\nonumber\\ && \hspace{3ex}
-\frac{1}{9} \left(\langle v^2_S \rangle_V \right)^2
- \frac{1}{12} \langle v^4_S \rangle_V
+ \langle v^2_D \rangle_V 
+ \frac{1}{8} \langle DE_1 \rangle_V
\bigg] \delta(x-1/2), 
\\
f_V^\perp (\mu_0) 
\phi_V^\perp{}^{(\alpha_s)} (x,\mu_0) &=&
f_V^{\parallel} \phi_V^\perp{}^{(\alpha_s)} (x,\mu_0), 
\\
f_V^\perp (\mu_0) 
\phi_V^\perp{}^{(2)} (x,\mu_0) &=&
f_V^{\parallel} \bigg[ 
\left( \frac{1}{3} - \frac{m_V - 2 m}{4 m} \right)
\langle v^2_S \rangle_V
+ \frac{7}{18} \left(\langle v^2_S \rangle_V \right)^2 
- \frac{11}{20} \langle v_S^4 \rangle_V 
\nonumber \\ && \hspace{5ex}
-\frac{1}{5} \langle v_D^2 \rangle_V 
- \frac{1}{8} \langle DE_1 \rangle
\bigg]
\frac{\delta^{(2)} (x-1/2)}{8},
\\
f_V^\perp (\mu_0) 
\phi_V^\perp{}^{(4)} (x,\mu_0) &=&
f_V^\parallel \frac{ \langle v_S^4 \rangle_V}{5} 
\frac{\delta^{(4)} (x-1/2)}{384}.
\end{eqnarray}
\end{subequations}
We use the results in Eqs.~(\ref{eq:LCDAresult3}) to compute the direct
amplitude. 
Similarly, when calculating the convolution in Eq.~(\ref{eq:ampdir}), 
we ignore the order $\alpha_s$ correction to $T_H (x, \mu)$ for $n=\alpha_s$,
2, and~4. 

We first discuss the numerical input parameters necessary for the 
 computation of the direct amplitude. 
We compute the decay constant $f_V^\parallel$ from 
Eq.~(\ref{eq:fvpar_from_rate}) using the measured leptonic decay rate, 
and $\alpha(\mu_0) = 1/132$ regardless of the vector meson state $V$. 
We set the central values of $\mu_0$ and $\mu$ to be $m_V$ and $m_H$,
respectively. 
We compute $\overline{m} (\mu)$, $\alpha_s(\mu_0)$ and $\alpha_s(\mu)$
using {\tt RunDec}~\cite{Chetyrkin:2000yt, Schmidt:2012az, Herren:2017osy}. 
We take the QED coupling constant at scale $\mu$ 
to be $\alpha (\mu) = 1/128$. 
We resum the logarithms in $\mu/\mu_0$ to NLL accuracy 
using the Abel-Pad\'e method~\cite{Bodwin:2016edd, Bodwin:2017wdu}. 
We take the number of active quark flavors in the evolution kernel 
to be $n_f = 4$ and 5 for scales 
below and above $m_b$, respectively.  
The ratios $m^2 \langle v_S^2 \rangle_V$ for $V = J/\psi$ and $\Upsilon(nS)$ 
have been obtained from potential-model (Cornell potential)
calculations in Refs.~\cite{Bodwin:2007fz, Chung:2010vz}: 
\begin{subequations}
\label{eq:vsq}%
\begin{eqnarray}
\frac{ \langle J/\psi | \psi^\dag \bm{\sigma} \cdot \bm{\epsilon} (\lambda)
(-\tfrac{i}{2} \overleftrightarrow{\bm{D}})^2 \chi | 0 \rangle }
{\langle J/\psi | \psi^\dag \bm{\sigma} \cdot \bm{\epsilon} (\lambda) \chi 
| 0 \rangle } 
&=& 0.441 {}^{+0.045}_{-0.046} \pm 0.132\textrm{~GeV}^2, 
\\
\frac{\langle \Upsilon(1S) | \psi^\dag \bm{\sigma} \cdot \bm{\epsilon} (\lambda)
(-\tfrac{i}{2} \overleftrightarrow{\bm{D}})^2 \chi | 0 \rangle }
{\langle \Upsilon(1S) | \psi^\dag \bm{\sigma} \cdot \bm{\epsilon} (\lambda) 
\chi | 0 \rangle } 
&=& 
-0.193 
{}^{+0.069}_{-0.070}
\pm 0.019
\textrm{~GeV}^2, 
\\
\frac{\langle \Upsilon(2S) | \psi^\dag \bm{\sigma} \cdot \bm{\epsilon} (\lambda)
(-\tfrac{i}{2} \overleftrightarrow{\bm{D}})^2 \chi | 0 \rangle }
{\langle \Upsilon(2S) | \psi^\dag \bm{\sigma} \cdot \bm{\epsilon} (\lambda) 
\chi | 0 \rangle } 
&=& 1.898 {}^{+0.090}_{-0.089} \pm 0.190\textrm{~GeV}^2, 
\\
\frac{\langle \Upsilon(3S) | \psi^\dag \bm{\sigma} \cdot \bm{\epsilon} (\lambda)
(-\tfrac{i}{2} \overleftrightarrow{\bm{D}})^2 \chi | 0 \rangle }
{\langle \Upsilon(3S) | \psi^\dag \bm{\sigma} \cdot \bm{\epsilon} (\lambda) 
\chi | 0 \rangle } 
&=& 3.283 {}^{+0.130}_{-0.127} \pm 0.328 \textrm{~GeV}^2.
\end{eqnarray}
\end{subequations}
The first uncertainty comes from the variation of the potential-model
parameters, and the second uncertainty is taken to be 
$\pm0.3$ and $\pm0.1$ times the
central value for $V = J/\psi$ and $V=\Upsilon(nS)$ respectively, 
which comes from the uncalculated corrections of relative order $v^2$. 
The potential-model calculations in 
Refs.~\cite{Bodwin:2007fz, Chung:2010vz} also let us compute the binding
energies $E_V = m_V - 2 m$, which are given by 
\begin{subequations}
\label{eq:binding}%
\begin{eqnarray}
E_{J/\psi} &=& 0.306 {}^{+0.039}_{-0.041} \pm 0.092\textrm{~GeV}, 
\\
E_{\Upsilon(1S)} &=& -0.053 \pm0.018 \pm 0.005 \textrm{~GeV}, 
\\
E_{\Upsilon(2S)} &=& 0.482 \pm0.032 \pm 0.048\textrm{~GeV}, 
\\
E_{\Upsilon(3S)} &=& 0.823 \pm 0.045 \pm 0.082\textrm{~GeV}, 
\end{eqnarray}
\end{subequations}
where the uncertainties are as in Eqs.~(\ref{eq:vsq}). 
The uncertainty in $E_V$ from the variation of the potential-model 
parameters are 
correlated with the uncertainty in the ratio $m^2 \langle v_S^2 \rangle_V$. 
We use these values of the binding energies to compute the heavy-quark mass
through the relation $m = \frac{1}{2} (m_V - E_V)$. 
The authors of Refs.~\cite{Bodwin:2007fz, Bodwin:2006dn} 
also found the relation 
$\langle v_S^4 \rangle_V = \left( \langle v_S^2 \rangle_V \right)^2
[1+O(v^2)]$, which is valid if the LDMEs are computed in a potential model 
like the Cornell potential and in dimensional regularization.
By using this relation, we take the central value for the ratio 
$\langle v_S^4 \rangle_V$ 
to be $\left( \langle v_S^2 \rangle_V \right)^2$ and take the uncertainty 
to be $\pm 0.3$ and $\pm 0.1$ times the central value 
of $\langle v_S^4 \rangle_V$ 
for $V = J/\psi$ and $V=\Upsilon(nS)$, respectively. 
The ratios $\langle v_D^2 \rangle_V$ and $\langle DE_1 \rangle_V$ are not 
known; since these ratios scale as $v^4$, we take the
central values of the ratios to be 0 and take the uncertainties to be 
$\pm 0.09$ for $V=J/\psi$, and $\pm 0.01$ for $V=\Upsilon(nS)$, respectively. 

We now list the sources of uncertainties in ${\cal A}_{\rm dir}$. 
We account for the uncertainties in the NRQCD LDMEs as discussed above. 
We vary the scales $\mu$ and $\mu_0$ between 
$\frac{1}{2} m_H < \mu < 2 m_H$ and $\frac{1}{2} m_V < \mu_0 < 2 m_V$, while 
we ignore the negligibly small shifts in the QED couplings $\alpha(\mu_0)$ and 
$\alpha(\mu)$ from scale variations. We also ignore the uncertainties from
$m_H$ and $m_V$. We consider the
 uncertainty in
$f_V^\parallel$ that originates from the experimental uncertainties in the
leptonic decay rate. We add the uncertainties in quadrature. 
Our numerical results for ${\cal A}_{\rm dir}$ are shown in
Table.~\ref{table:amplitudes}. 
The imaginary part of ${\cal A}_{\rm dir}$ comes from the imaginary part in the
order $\alpha_s$ correction to $T_H(x,\mu)$. 
Note that the uncertainties in the real and imaginary parts of 
${\cal A}_{\rm dir}$ are correlated. 

For $J/\psi$, the uncertainties from the LDMEs are comparable to the uncertainties
from the variations of $\mu_0$ and $\mu$. 
If we ignore the uncertainties from scale variations, the uncertainty in 
${\rm Re} [{\cal A}_{\rm dir}]$ is about 8\% of the central value, and 
the uncertainty in ${\rm Im} [{\cal A}_{\rm dir}]$ is about 10\% of the central
value. This is comparable to the nominal size of the relative order $v^4$
correction. 

For $\Upsilon(nS)$, the uncertainties are dominated by scale variations. 
If we ignore the uncertainties from scale variations, the uncertainties in the
real and imaginary parts of ${\cal A}_{\rm dir}$ are about 1\% of the
central values, which are comparable to the nominal size of the relative
order $v^4$ correction. 

While the uncertainties from the LDMEs are expected to be reduced when the
LDMEs $\langle v_D^2 \rangle_V$ and $\langle DE_1 \rangle_V$ are constrained,
the reduction of the uncertainties from variations of scales would require
calculation of the order $\alpha_s^2$ and order $\alpha_s v^2$ correction to
the decay constant and the LCDA, and the order $\alpha_s^2$ correction to 
$T_H (x,\mu)$. 

We again compare our results for ${\cal A}_{\rm dir}$ with 
the
calculation in Ref.~\cite{Bodwin:2017wdu}. 
Our results for ${\cal A}_{\rm dir}$ in Table.~\ref{table:amplitudes}
are compatible with those in Ref.~\cite{Bodwin:2017wdu} within uncertainties. 
For $J/\psi$, the uncertainty for ${\cal A}_{\rm dir}$ is smaller 
than in Ref.~\cite{Bodwin:2017wdu}, owing to the explicit calculation of the
relative order $v^4$ corrections included in this work. 
On the other hand, for $\Upsilon(nS)$, the uncertainties for 
${\cal A}_{\rm dir}$ are larger than those in Ref.~\cite{Bodwin:2017wdu}; 
the main reason is that in Ref.~\cite{Bodwin:2017wdu}, the uncertainty from
variations in the scales $\mu_0$ and $\mu$ were not taken into account, 
and instead, the uncertainties from 
the uncalculated order $\alpha_s^2$ and order $\alpha_s v^2$ corrections to the
real part of ${\cal A}_{\rm dir}$ were 
estimated to be $C_F C_A \alpha_s^2 (m)/\pi^2$ and $C_F \alpha_s(m) v^2/\pi$ of
the central value, 
and the uncertainty from the uncalculated order $\alpha_s^2$ correction to the
imaginary part was estimated to be $C_A \alpha_s(m)/\pi$ of the central value. 
We note that these estimates lead to smaller uncertainties compared to 
uncertainties estimated from variations of the scales $\mu_0$ and $\mu$. 
If we would use the same uncertainty estimates used in Ref.~\cite{Bodwin:2017wdu} for
$\Upsilon(nS)$, then the uncertainties in ${\rm Re} [{\cal A}_{\rm dir}]$ 
would reduce by a factor of 2, while the uncertainties in 
${\rm Im} [{\cal A}_{\rm dir}]$ would increase slightly.

\subsection{Decay rate} 

\begin{table}[t]
\begin{center}
\begin{ruledtabular}
\begin{tabular}{ccc}
$V$ & 
${\cal A}_{\rm ind} \times 10^5$~(GeV${}^{1/2}$) & 
${\cal A}_{\rm dir} \times 10^5$~(GeV${}^{1/2}$) 
\\
\hline
$J/\psi$ & $- (11.73^{+0.16}_{-0.16})$ & 
$\phantom{-}(0.631^{+0.071}_{-0.080}) + (0.065^{+0.015}_{-0.012}) i$
\\
$\Upsilon(1S)$ & $\phantom{+} (3.288^{+0.033}_{-0.033})$ & 
$-(2.719^{+0.136}_{-0.142}) - (0.291^{+0.055}_{-0.040}) i$
\\
$\Upsilon(2S)$ & $\phantom{+} (2.158^{+0.026}_{-0.026})$ & 
$-(1.896^{+0.101}_{-0.104}) - (0.197^{+0.037}_{-0.027}) i$
\\
$\Upsilon(3S)$ & $\phantom{+} (1.808^{+0.022}_{-0.022})$ & 
$-(1.614^{+0.090}_{-0.093}) - (0.164^{+0.031}_{-0.023}) i$
\\
\end{tabular}
\end{ruledtabular}
\caption{Numerical results for the
amplitudes 
${\cal A}_{\rm ind}$ and
${\cal A}_{\rm dir}$.
\label{table:amplitudes}%
}
\end{center}
\end{table}

We now compute the total decay rate from the numerical results of
${\cal A}_{\rm ind}$ and ${\cal A}_{\rm dir}$ in Table.~\ref{table:amplitudes}.
Our results for the decay rates $\Gamma( H \to V + \gamma)$ are shown in 
Table~\ref{table:rates}. 
When computing the uncertainties in $\Gamma( H \to V + \gamma)$, we consider 
the correlation between the uncertainties in the real and imaginary parts of 
${\cal A}_{\rm dir}$. We also consider the correlation between the
uncertainties in ${\cal A}_{\rm ind}$ and ${\cal A}_{\rm dir}$ that comes from
the measured leptonic decay rates $\Gamma(V \to \ell^+ \ell^-)$. The
uncertainty from 
uncalculated corrections of relative order $m_V^2/m_H^2$ is taken to be
$m_V^2/m_H^2$ of the central value of the decay rate. 
We also compute the branching ratios ${\rm Br} (H \to V + \gamma)$ by using 
the total decay rate of the Higgs $\Gamma_H$ computed in
Refs.~\cite{Dittmaier:2011ti, Dittmaier:2012vm}: 
for $m_H = 125.18$~GeV, it is $\Gamma_H = 4.10$~MeV, with uncertainties
given by $+4.0$\% and $-3.9$\% of the central value. 
Our results for the branching ratios ${\rm Br} (H \to V + \gamma)$ are shown in 
Table~\ref{table:rates}.

\begin{table}[t]
\begin{center}
\begin{ruledtabular}
\begin{tabular}{ccc}
$V$ &
$\Gamma(H\to V + \gamma)$~(GeV) &
${\rm Br} (H \to V + \gamma)$
\\
\hline
$J/\psi$ & $(1.231^{+0.038}_{-0.037}) \times 10^{-8}$ &
$(3.01^{+0.15}_{-0.15}) \times 10^{-6}$
\\
$\Upsilon(1S)$ & $\phantom{+} (4.08{}^{+1.65}_{-1.23}) \times 10^{-11}$ &
$(9.97 ^{+4.04}_{-3.03}) \times 10^{-9}$
\\
$\Upsilon(2S)$ & $\phantom{+} (1.07{}^{+0.57}_{-0.37}) \times 10^{-11}$ &
$(2.62 ^{+1.39}_{-0.91}) \times 10^{-9}$
\\
$\Upsilon(3S)$ & $\phantom{+} (0.77{}^{+0.43}_{-0.28}) \times 10^{-11}$ &
$(1.87^{+1.05}_{-0.69}) \times 10^{-9}$
\\
\end{tabular}
\end{ruledtabular}
\caption{Numerical results for 
$\Gamma(H \to V + \gamma)$ 
and ${\rm Br} (H \to V + \gamma)$. 
\label{table:rates}%
}
\end{center}
\end{table}

For $V=J/\psi$, ${\cal A}_{\rm ind}$ is more than an order of magnitude larger
than ${\cal A}_{\rm dir}$, and so, the uncertainty in 
the decay rate $\Gamma( H \to J/\psi + \gamma)$ is dominated by
the uncertainty in ${\cal A}_{\rm ind}$. 
As a result, the uncertainty in the prediction for $\Gamma( H \to J/\psi +
\gamma)$ is about 3\% of the central value. 
On the other hand, for $V=\Upsilon (nS)$, ${\cal A}_{\rm ind}$ and 
${\cal A}_{\rm dir}$ are comparable. Due to the large cancellation
between ${\cal A}_{\rm ind}$ and ${\cal A}_{\rm dir}$ for $\Upsilon(nS)$, 
the uncertainty in $\Gamma[ H \to \Upsilon(nS) + \gamma]$ is sensitive to the
uncertainty in ${\cal A}_{\rm dir}$. 

Our results are compatible with the previous calculation in
Ref.~\cite{Bodwin:2017wdu} within errors. 
The uncertainties in $\Gamma( H \to J/\psi + \gamma)$ and 
${\rm Br} (H \to J/\psi + \gamma)$ are slightly smaller than those of 
Ref.~\cite{Bodwin:2017wdu}, which is a result of the reduction of the 
uncertainty in ${\cal A}_{\rm ind}$ resulting from the 
improved measurement of $m_H$. 
On the other hand, the uncertainties in $\Gamma[ H \to \Upsilon(nS) + \gamma]$ 
and ${\rm Br} [H \to \Upsilon(nS) + \gamma]$ are slightly larger than those of
Ref.~\cite{Bodwin:2017wdu}. As we discussed in the previous section, 
if we would use the same estimates for the uncertainties used in
Ref.~\cite{Bodwin:2017wdu}, then the uncertainties in 
${\cal A}_{\rm dir}$ would be 
reduced, leading to uncertainties in 
$\Gamma[ H \to \Upsilon(nS) + \gamma]$ 
and ${\rm Br} [H \to \Upsilon(nS) + \gamma]$ that are  smaller than 
those of Ref.~\cite{Bodwin:2017wdu}. 

\begin{figure}
\epsfig{file=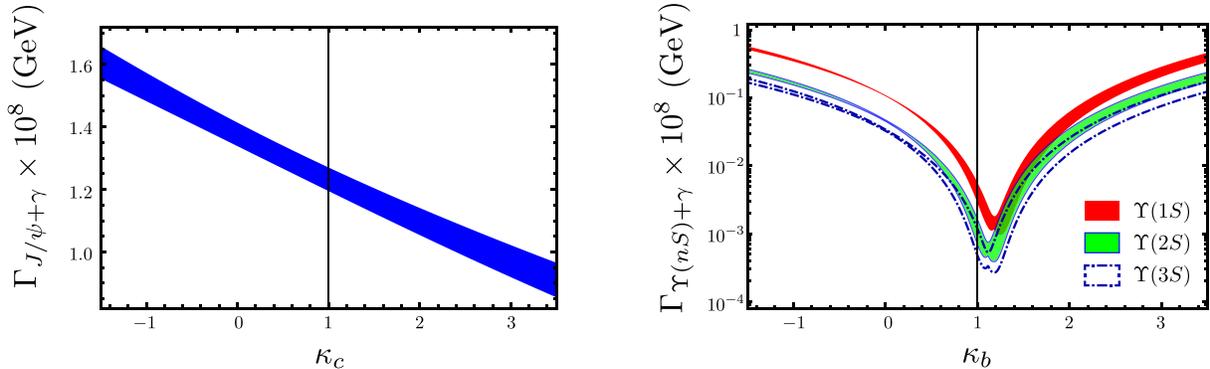,width=16cm}
\caption{
\label{fig:higgs}%
The decay rate $\Gamma_{J/\psi+\gamma} = 
\Gamma (H \to J/\psi + \gamma)$ 
when the Higgs charm coupling is rescaled by a factor of $\kappa_c$, 
and the decay rates 
$\Gamma_{\Upsilon(nS)+\gamma} = 
\Gamma [H \to \Upsilon(nS) + \gamma]$ for $n = 1,2$, and 3 
when the Higgs bottom coupling is rescaled by a factor of $\kappa_b$. 
The Standard Model result corresponds to $\kappa_c = 1$ and $\kappa_b =
1$.  }
\end{figure}

We also consider the case where the Yukawa coupling $y_Q$ deviates from the
Standard Model by a factor of $\kappa_Q$. 
In this case, the Higgs decay rate into $V + \gamma$ is given by 
$| \kappa_Q {\cal A}_{\rm dir} + {\cal A}_{\rm ind} | ^2$. 
The decay rates 
$\Gamma[H \to J/\psi + \gamma]$ and $\Gamma[ H \to \Upsilon(nS) +
\gamma]$ for $-1.5<\kappa_Q<3.5$ are plotted in Fig.~\ref{fig:higgs}.
For $J/\psi$, the decay rate shows moderate dependence on $\kappa_c$, while for
$\Upsilon(nS)$, the decay rates are very sensitive to $\kappa_b$. 
In the case of $J/\psi$, a reduction of uncertainties in ${\cal A}_{\rm dir}$
will only have a small effect on the sensitivity of the $H \to J/\psi + \gamma$ 
rate on $\kappa_c$, because the uncertainty in 
$\Gamma( H \to J/\psi + \gamma)$ is dominated by the
uncertainty in ${\cal A}_{\rm ind}$. 
On the other hand, for $\Upsilon$, the decay rate $\Gamma[ H \to \Upsilon(nS) +
\gamma]$ will be even more sensitive to $\kappa_b$ if the accuracy in 
${\cal A}_{\rm dir}$ is improved. The reason for this sensitivity is the
large cancellation between ${\cal A}_{\rm dir}$ and ${\cal A}_{\rm ind}$ that
only happens for $\kappa_b$ close to~1.
We note that even though the Standard-Model values of 
$\Gamma[ H \to \Upsilon(nS) + \gamma]$ are very small, 
the decay rate $\Gamma[H \to \Upsilon(nS) + \gamma]$ can still probe large
deviations of the Higgs bottom coupling from its Standard Model value. Indeed 
$\Gamma[H \to \Upsilon(nS) + \gamma]$ may be two/three orders of magnitude
larger than the Standard Model value for $\kappa_b \lesssim -1$ or $\kappa_b
\gtrsim 3$. In particular, the case $\kappa_b \lesssim -1$ corresponds to a
Higgs bottom coupling whose sign is opposite to the Standard Model Yukawa
coupling.

\section{Summary and Discussion} 
\label{sec:summary}%

In this work we computed the order $v^4$ correction to the decay rate 
$\Gamma(H \to V + \gamma)$ where $V = J/\psi$ or $\Upsilon(nS)$ based on the
nonrelativistic QCD (NRQCD) factorization formalism. 
By using the light-cone approach, we resummed the logarithms of $m_H^2/m_V^2$
that appear in higher order corrections in $\alpha_s$ to all orders in
$\alpha_s$ at the next-to-leading logarithmic accuracy. 

If we consider that $\alpha_s \approx v^2$ at the scale of the heavy-quark
mass, the corrections of order $\alpha_s^2$ and $\alpha_s v^2$ would be of the
same order as the order $v^4$ corrections computed in this paper. The
calculation of order $\alpha_s^2$ and order $\alpha_s v^2$ corrections to the
decay rate $\Gamma(H \to V + \gamma)$ requires calculation of the two-loop
correction to $c_0$ and the one-loop correction to $c_{\bm{D}^2}$ 
in Eq.~(\ref{eq:Jpsifactorization}), respectively. 
In the light-cone approach, the order $\alpha_s^2$ correction to the hard part 
$T_H(x,\mu)$ in Eq.~(\ref{eq:TH}), as well as the two-loop correction to 
$\tilde{c}_0(x)$ and the one-loop correction to $\tilde{c}_{\bm{D}^2} (x)$ 
are necessary. Since these corrections have not been computed, we include the
effects of these higher-order corrections in the uncertainties.

Our numerical results for the Standard Model values of 
the decay rates $\Gamma(H \to V + \gamma)$ and branching ratios 
${\rm Br} (H \to V + \gamma)$ are shown in Table~\ref{table:rates}. 
The corrections computed in this work 
improve the theoretical accuracy in the prediction of the decay rate
$\Gamma(H \to J/\psi + \gamma)$  
compared to a previous calculation in Ref.~\cite{Bodwin:2017wdu}. 
If we would have used the same method to estimate the uncertainties as in
Ref.~\cite{Bodwin:2017wdu}, we would have found 
uncertainties in $\Gamma[H \to \Upsilon(nS) + \gamma]$ 
reduced by almost a factor of two compared
to our results in Table~\ref{table:rates}, so that they would
become comparable to the
results in Ref.~\cite{Bodwin:2017wdu}, as the error of 
$\Gamma[H \to \Upsilon(nS) + \gamma]$ would
be dominated by the scale uncertainties.
However, due to the difference in the method to estimate uncertainties, our
results for $\Gamma[H \to \Upsilon(nS) + \gamma]$ have larger uncertainties
than those of Ref.~\cite{Bodwin:2017wdu}. 

We note that the decay rates $\Gamma(H \to V + \gamma)$
depend on the LDMEs $\langle V | \psi^\dag 
\epsilon^i (\lambda) \sigma^j (-\tfrac{i}{2})^2 $ $\times
\overleftrightarrow{\bm{D}}^{(i} \overleftrightarrow{\bm{D}}^{j)} 
\chi | 0 \rangle$ and
$\langle V | \psi^\dag \bm{\epsilon} (\lambda)  \cdot 
\tfrac{1}{2} 
[\bm{\sigma} \times ( \overleftrightarrow{\bm{D}} \times g_s \bm{E}
- g_s \bm{E} \times \overleftrightarrow{\bm{D}} )] \chi | 0 \rangle$ that are
currently unknown.
In our numerical results, we have estimated their sizes 
according to the conservative power counting in 
Refs.~\cite{Brambilla:2001xy, Brambilla:2002nu, Brambilla:2006ph,
Brambilla:2008zg}, and included
their effects in the uncertainties. 
For $V = J/\psi$, the uncertainties from these unknown LDMEs are significant
compared to the total uncertainty in $\Gamma(H \to J/\psi + \gamma)$. 
Therefore, knowledge of these LDMEs will improve the accuracy in the prediction
of the decay rate $\Gamma(H \to J/\psi + \gamma)$. 
Also the systematic inclusion of relative order $v^2$ effects in the
determination of the matrix element 
$\langle V | \psi^\dag \bm{\sigma} \cdot \bm{\epsilon} (\lambda)
(-\tfrac{i}{2} \overleftrightarrow{\bm{D}})^2 \chi | 0 \rangle $
would significantly improve the determination of the $H \to J/\psi + \gamma$
decay rate. 
A calculation of the LDMEs in potential NRQCD may help
constrain these LDMEs~\cite{Brambilla:2002nu}.

In Ref.~\cite{Bodwin:2013gca}, the authors estimated that the process 
$H \to J/\psi + \gamma$ could
be measured at the HL-LHC experiment through the leptonic decays 
of $J/\psi$ into $e^+ e^-$ and $\mu^+ \mu^-$.
However, a more recent study by the ATLAS Collaboration found that the
HL-LHC would only be able to put an upper bound for the decay rate 
$\Gamma (H \to J/\psi + \gamma)$ at 95\% confidence level 
that is about 15 times larger the Standard Model
value~\cite{ATL-PHYS-PUB-2015-043}. 
Although the prospect for measuring the process $H \to J/\psi + \gamma$ at the
HL-LHC does not look so good at the moment, it is possible
that the experimental methods will 
improve over time; it is also possible that future experiments at the International
Linear Collider, 
the Circular Electron Positron Collider,
the Compact Linear Collider, or the Future Circular Collider
 will be able to probe
such processes. 
In the case of the $\Upsilon(nS)$, the Standard Model values for the decay
rates $\Gamma [H \to \Upsilon(nS) + \gamma]$ are about three orders of magnitude
smaller than $\Gamma (H \to J/\psi + \gamma)$, owing to large cancellations 
between the direct and indirect amplitudes. 

We show the decay rates $\Gamma(H \to J/\psi + \gamma)$ 
and $\Gamma[H \to \Upsilon(nS) + \gamma]$ 
when the Higgs charm coupling is rescaled by
a factor $\kappa_c$ and the 
Higgs bottom coupling is rescaled by
a factor $\kappa_b$ compared to the Standard Model in 
Fig.~\ref{fig:higgs}.
Due to the cancellation between direct and indirect amplitudes that
occurs when $\kappa_b \approx 1$, the decay
rates $\Gamma[H \to \Upsilon(nS) + \gamma]$ are highly sensitive to $\kappa_b$,
so much so that the combined rate of 
$\Gamma [H \to \Upsilon(nS) + \gamma]$ for $n=1,2$, and $3$ 
increases to be larger than one half of the 
Standard Model value of $\Gamma(H \to J/\psi + \gamma)$
for $\kappa_b \lesssim -1$ or $\kappa_b \gtrsim 3$. Therefore,
the numerical results presented in this paper 
may be useful in determining the size and sign of the 
Higgs charm and even more the Higgs bottom couplings when these processes are measured in
future experiments.

\begin{acknowledgments}

We thank Geoffrey Bodwin for helpful discussions. 
The research of N.~B.\ is supported  by the DFG Grant No.~BR~4058/2-2.
N.~B.\ and A.~V.\ acknowledge support from the DFG cluster of excellence
``Origins'' (www.origins-cluster.de).
The work of H.~S.~C.\ is supported by the Alexander von Humboldt Foundation.
The work of V.~S.\ is supported in part by the National Science
Foundation of China (No.~11135006, No.~11275168, No.~11422544, No.~11375151, 
No.~11535002)
and the Zhejiang University Fundamental Research Funds for the Central
Universities (2017QNA3007).

\end{acknowledgments}

\appendix

\section{Kinematics}
\label{sec:kinematics}%

In this section, we present the kinematical conventions that we use for the
perturbative $Q \bar Q$ and the $Q \bar Q g$ states. 
The kinematical conventions that we use here are identical to the ones used in 
Ref.~\cite{Brambilla:2017kgw}. 

\subsection{Two-body kinematics}

We let the $Q$ and the $\bar Q$ to have the momenta $p_1$ and $p_2$,
respectively. 
We denote the total momentum of the $Q \bar Q$ system as $P = p_1 + p_2$, 
and the relative momentum of the $Q$ and the $\bar Q$ as 
$q = \frac{1}{2} (p_1 - p_2)$. 
In the rest frame of the $Q \bar Q$, $q$ and $P$ are 
given by $q = (0,\bm{q})$, and $P = (2 \sqrt{m^2+\bm{q}^2}, \bm{0})$, 
which leads to $p_1 = (\sqrt{m^2+\bm{q}^2}, \bm{q})$ and 
$p_2 = (\sqrt{m^2+\bm{q}^2}, -\bm{q})$.

\subsection{Three-body kinematics}

We again let the $Q$ and the $\bar Q$ to have the momenta $p_1$ and $p_2$,
respectively, and the gluon carry momentum $k_g$. 
We set the total momentum of the $Q \bar Q g$ system to be $P = p_1 + p_2 +
k_g$. 
We define $q_1 = \frac{1}{2} (p_1-p_2)$ and $q_2 = \frac{1}{6} (2 k_g - p_1 -
p_2)$, so that 
$p_1 = \frac{1}{3} P + q_1 - q_2$, 
$p_2 = \frac{1}{3} P - q_1 - q_2$, 
and 
$k_g = \frac{1}{3} P + 2 q_2$. 
In the rest frame of the $Q \bar Q g$ system, where $\bm{P} = \bm{p}_1 +
\bm{p}_2 + \bm{k}_g = 0$, 
$p_1 = (\sqrt{m^2 + (\bm{q}_1 - \bm{q}_2)^2},
\bm{q}_1 - \bm{q}_2)$, 
$p_2 = (\sqrt{m^2 + (\bm{q}_1 + \bm{q}_2)^2},
- \bm{q}_1 - \bm{q}_2)$, 
and
$k_g = (2 | \bm{q}_2| , 2 \bm{q}_2)$, so that 
$P^0 = 2 |\bm{q}_2| + \sqrt{(\bm{q}_1 + \bm{q}_2)^2+m^2}
+ \sqrt{(\bm{q}_1 - \bm{q}_2)^2+m^2}$. 

\section{Nonrelativistic expansion of Dirac spinors}
\label{sec:spinprj}%

In order to obtain the matching conditions, we need to express the QCD
amplitudes in terms of the 3-momenta of the particles and the 2-component Pauli
spinors in the frame where the NRQCD LDMEs are defined. 
One way to accomplish this is to compute the QCD amplitudes using explicit 
forms of the Dirac spinors $u(p_1,s)$ and $v(p_2,s')$ with definite spin. 
In this appendix, we introduce a simple way to compute the QCD amplitudes with 
the explicit Dirac spinors that can be easily used in automated calculations. 

For the perturbative $Q \bar Q$ or the $Q \bar Q g$ states, the frame where the
NRQCD LDMEs are defined is 
the frame where the total momentum $P$ of the state is at rest 
($\bm{P} = \bm{0}$), so that $P^0 = \sqrt{P^2}$ and 
$A^0 = A \cdot P/\sqrt{P^2}$ for an arbitrary 4-vector $A$. 
In this frame, 
the nonrelativistically normalized Dirac spinors in Dirac basis are given
by~\cite{Bodwin:1994jh} 
\begin{eqnarray}
u(p_1,s) &=& 
N_1 
\begin{pmatrix} (E_1 + m_1) \xi_s \\ 
\bm{\sigma} \cdot \bm{p}_1 \xi_s \end{pmatrix} 
=
 N_1 (p\!\!\!/_1 + m_1) 
\begin{pmatrix} \xi_s \\ 0 \end{pmatrix}, 
\\
v(p_2,s') &=& 
N_2 
\begin{pmatrix} 
\bm{\sigma} \cdot \bm{p}_2 \eta_{s'} 
 \\ (E_2 + m_2) \eta_{s'} \end{pmatrix} 
=
- N_2 (p\!\!\!/_2 - m_2) 
\begin{pmatrix} 0 \\ \eta_{s'} \end{pmatrix}, 
\end{eqnarray}
where $m_i^2 = p_i^2$, $E_i = p_i \cdot P/\sqrt{P^2}$, and 
$N_i = 1/\sqrt{2 E_i (E_i + m_i)}$. In general, $m_1$ and $m_2$ can
be different. 
A QCD amplitude for production of a quark with momentum $p_1$ and an 
antiquark with momentum $p_2$ involves 
\begin{equation}
v(p_2,s') \otimes \bar u(p_1,s)
=
- N_1 N_2
(p\!\!\!/_2 - m_2) 
\begin{pmatrix} 0 & 0 \\
\eta_{s'} \otimes \xi_s^\dag & 0 
\end{pmatrix} 
(p\!\!\!/_1 + m_1).
\end{equation}
For example, $\bar u (p_1,s) \Gamma v (p_2,s') = 
{\rm tr} [ \Gamma v(p_2,s') \otimes \bar u(p_1,s)]$, where $\Gamma$ is a
product of gamma matrices.
Since $\eta_{s'} \otimes \xi_s^\dag$ is a $2 \times 2$ matrix, 
it can be written as 
a linear combination of a $2 \times 2$ identity matrix $I$ and the 
$\bm{\sigma}$ 
matrices:
\begin{equation}
\eta_{s'} \otimes \xi_s^\dag
= \frac{1}{2} {\rm tr} ( \eta_{s'} \otimes \xi_s^\dag) I
+ \frac{1}{2} {\rm tr} ( \eta_{s'} \otimes \xi_s^\dag \bm{\sigma}) 
\cdot \bm{\sigma}
= \frac{1}{2} \xi_s^\dag \eta_{s'} I
+ \frac{1}{2} (\xi_s^\dag \bm{\sigma} \eta_{s'}) \cdot \bm{\sigma} . 
\end{equation}
We note that in the Dirac basis, 
\begin{equation}
\begin{pmatrix} 0 & 0 
\\ \bm{\sigma} & 0 \end{pmatrix}
= 
\begin{pmatrix} 0 & - \bm{\sigma}
\\ \bm{\sigma} & 0 \end{pmatrix}
\begin{pmatrix} 1 & 0
\\ 0 & 0 \end{pmatrix}
= -\bm{\gamma} \frac{\gamma_0 + 1}{2}
= -\bm{\gamma} \frac{P\!\!\!/ + \sqrt{P^2}}{2 \sqrt{P^2}}, 
\end{equation}
and
\begin{equation}
\begin{pmatrix} 0 & 0 
\\ 1 & 0 \end{pmatrix}
= 
\begin{pmatrix} 0 & 1
\\ 1 & 0 \end{pmatrix}
\begin{pmatrix} 1 & 0
\\ 0 & 0 \end{pmatrix}
= \gamma_5 \frac{\gamma_0 + 1}{2}
= \gamma_5 \frac{P\!\!\!/ + \sqrt{P^2}}{2 \sqrt{P^2}}, 
\end{equation}
so that 
\begin{equation}
\label{eq:project}%
v(p_2,s') \otimes \bar u(p_1,s)
=
- \frac{N_1 N_2}{4 \sqrt{P^2}}
(p\!\!\!/_2 - m_2) 
\left[
(\xi_s^\dag \eta_{s'}) \gamma_5
- (\xi_s^\dag \bm{\sigma} \eta_{s'}) \cdot \bm{\gamma} \right]
\left(P\!\!\!/ + \sqrt{P^2} \right)
(p\!\!\!/_1 + m_1).
\end{equation}
By using Eq.~(\ref{eq:project}), we can compute the quantities of the form 
$\bar u (p_1,s) \Gamma v (p_2,s') = 
{\rm tr} [ \Gamma v(p_2,s') \otimes \bar u(p_1,s)]$, where $\Gamma$ is any 
product of gamma matrices, as traces of gamma matrices. 
This can be easily implemented in automated calculations using 
{\tt FeynCalc}.
We obtain the nonrelativistic expansion of a QCD amplitude by computing the
amplitude using Eq.~(\ref{eq:project}) and expanding in powers of the
small 3-momenta of the $Q \bar Q$ and the $Q \bar Q g$ states. 
Using this method, we easily reproduce the explicit expressions for 
$\bar u (p_1,s) \Gamma v (p_2,s')$ for 
$\Gamma = 1$, $\gamma^\mu$, 
$\gamma^\mu \gamma^\nu - \gamma^\nu \gamma^\mu$, 
and 
$\gamma^\mu \gamma^\nu \gamma^\sigma - \gamma^\sigma \gamma^\nu \gamma^\mu$ 
found in Ref.~\cite{Braaten:1996rp}. 
We use this method to compute the QCD amplitudes in 
Eqs.~(\ref{eq:QQamplitude}), (\ref{eq:QQgamplitude}), 
(\ref{eq:LCDAQQ}), and (\ref{eq:LCDAQQg}). 

The expression in Eq.~(\ref{eq:project}) 
may serve to relate the
nonrelativistic expansion method that we use in this work with the covariant 
spin-projector method used in previous calculations of the $H \to J/\psi +
\gamma$ process in Refs.~\cite{Bodwin:2013gca, Bodwin:2014bpa}. 
In the covariant spin-projector method, spin-singlet and 
spin-triplet contributions are computed separately.
The standard forms of the spin projectors, such as the ones used in 
Ref.~\cite{Bodwin:2002hg}, can be obtained from Eq.~(\ref{eq:project}) by
projecting to a spin-singlet or a spin-triplet state using the Clebsch-Gordan
coefficients. For a spin-singlet state,
$\sum_{s,s'} \langle \tfrac{1}{2} s, \tfrac{1}{2} s' | 0 0 \rangle 
\xi_s^\dag \bm{\sigma} \eta_{s'}/\sqrt{2}$ vanishes, and 
$\sum_{s,s'} \langle \tfrac{1}{2} s, \tfrac{1}{2} s' | 0 0 \rangle 
\xi_s^\dag \eta_{s'}/\sqrt{2}$ is, up to a phase, equal to $1$. 
Hence, the spin-singlet projector is, up to a phase, given by the 
contribution in
Eq.~(\ref{eq:project}) that is proportional to $\xi^\dag_s \eta_{s'}$,
with $\xi^\dag_s \eta_{s'}$ replaced by $1$.
Similarly, for a spin-triplet state with polarization $\lambda$, 
$\sum_{s,s'} \langle \tfrac{1}{2} s, \tfrac{1}{2} s' | 1 \lambda \rangle 
\xi_s^\dag \eta_{s'}/\sqrt{2}$ vanishes for all $\lambda \in
\{-1,0,+1\}$, and 
$\sum_{s,s'} \langle \tfrac{1}{2} s, \tfrac{1}{2} s' | 1 \lambda \rangle 
\xi_s^\dag \bm{\sigma} \eta_{s'}/\sqrt{2}$ is, up to a phase, 
equal to the polarization vector of the spin-triplet state.
Therefore, the spin-triplet projector is, up to a phase, 
given by the contribution in 
Eq.~(\ref{eq:project}) that is proportional to 
$\xi^\dag_s \bm{\sigma} \eta_{s'}$, with 
$\xi^\dag_s \bm{\sigma} \eta_{s'}$ replaced by the polarization vector of the
spin-triplet state. 
As a result, the spin-singlet (spin-triplet) contribution of a QCD amplitude 
computed using the covariant spin-projector method is, up to a phase, 
equivalent to the contribution proportional to $\xi^\dag_s \eta_{s'}$ 
($\xi^\dag_s \bm{\sigma} \eta_{s'}$) in the amplitude computed in the 
nonrelativistic expansion method.
The phase conventions for the spin-singlet and the spin-triplet 
projectors depend on the conventions for the Clebsch--Gordan 
coefficients and the Pauli spinors $\xi$ and $\eta$. 

One advantage of using the covariant spin-projector method is that 
covariant expressions of the QCD amplitudes can be obtained easily, unlike the 
nonrelativistic expansion method in Ref.~\cite{Braaten:1996rp}. 
On the other hand, covariant expressions can also be obtained if we use 
Eq.~(\ref{eq:project}) to carry out the nonrelativistic expansion of spinors. 
Also, in this work, there is no advantage in computing the 
spin-singlet and spin-triplet contributions separately, 
because both contributions appear in the matching 
condition in Eq.~(\ref{eq:QQgfactorization}) simultaneously. 
Therefore, we compute the QCD amplitudes in 
Eqs.~(\ref{eq:QQamplitude}), (\ref{eq:QQgamplitude}), 
(\ref{eq:LCDAQQ}), and (\ref{eq:LCDAQQg}) using 
Eq.~(\ref{eq:project}). 

\section{Short-distance coefficients for $\bm{H\to h_c + \gamma}$}
\label{sec:axialvector}%

The calculation of the short-distance coefficients 
presented in Sec.~\ref{sec:fixedorder} can be easily applied to 
production amplitudes of other quarkonium states. 
For example, projecting onto the $J^{PC} = 1^{+-}$ state
gives us the short-distance coefficients for Higgs decay into $h_c + \gamma$ to
relative order $v^2$. To achieve this, we first write down the matching
condition for the $J^{PC} = 1^{+-}$ case as 
\begin{eqnarray}
&& \hspace{-5ex}
i {\cal M} [H \to Q \bar Q(J^{PC}=1^{+-}) + \gamma] 
\nonumber  \\ &=& 
\frac{c_1}{m} 
\langle Q \bar Q | \psi^\dag 
(-\tfrac{i}{2} \overleftrightarrow{\bm{D}}) \cdot \bm{\epsilon} (\lambda) 
\chi | 0 \rangle 
\nonumber  \\ && 
+ 
\frac{c_{\bm{D}^3}}{m^3} \langle Q \bar Q | \psi^\dag 
\bm{\epsilon} (\lambda) \cdot 
[-\tfrac{i}{2} \overleftrightarrow{\bm{D}}]^3
\chi | 0 \rangle 
+ O(g_s, (|\bm{q}|/m)^5),
\end{eqnarray}
and 
\begin{eqnarray}
&& \hspace{-5ex}
i {\cal M} [H \to Q \bar Q g (J^{PC}=1^{+-}) + \gamma ] 
\nonumber \\ 
&=& 
\frac{c_1}{m} 
\langle Q \bar Q g | \psi^\dag 
(-\tfrac{i}{2} \overleftrightarrow{\bm{D}}) \cdot \bm{\epsilon} (\lambda) 
\chi | 0 \rangle 
+ \frac{c_{\bm{D}^3}}{m^3} \langle Q \bar Q g | \psi^\dag 
\bm{\epsilon} (\lambda) \cdot 
[-\tfrac{i}{2} \overleftrightarrow{\bm{D}}]^3
\chi | 0 \rangle 
\nonumber \\ && 
+ \frac{c_E}{m^2} \langle Q \bar Q g | 
\psi^\dag g_s \bm{E} \cdot \bm{\epsilon} (\lambda) \chi | 0 \rangle 
\nonumber\\ && 
+ \frac{c_{DB_0}}{m^3} 
\langle Q \bar Q g | \psi^\dag 
\bm{\epsilon} (\lambda) \cdot \bm{\sigma}
\tfrac{1}{3} 
( \overleftrightarrow{\bm{D}} \cdot g_s \bm{B} 
+g_s \bm{B} \cdot \overleftrightarrow{\bm{D}} ) \chi | 0 \rangle 
\nonumber\\ && 
+ \frac{c_{DB_1}}{m^3} 
\langle Q \bar Q g | \psi^\dag 
\bm{\epsilon} (\lambda) \cdot 
\tfrac{1}{2} 
[\bm{\sigma} \times ( \overleftrightarrow{\bm{D}} \times g_s \bm{B} 
- g_s \bm{B} \times \overleftrightarrow{\bm{D}} )] \chi | 0 \rangle 
\nonumber\\ && 
+ \frac{c_{DB_1'}}{m^3} 
\langle Q \bar Q g | \psi^\dag 
\bm{\epsilon} (\lambda) \cdot 
\tfrac{i}{2} 
( \overleftrightarrow{\bm{D}} \times g_s \bm{B} 
+g_s \bm{B} \times \overleftrightarrow{\bm{D}} ) \chi | 0 \rangle 
\nonumber\\ && 
+ \frac{c_{DB_2}}{m^3} 
\langle Q \bar Q g | \psi^\dag 
\epsilon^i (\lambda) \sigma^j 
( \overleftrightarrow{\bm{D}}^{(i} g_s \bm{B}^{j)} 
+ g_s \bm{B}^{(i} \overleftrightarrow{\bm{D}}^{j)} )
\chi | 0 \rangle 
+ O(g_s^2, |\bm{q}_i|^3/m^3),
\end{eqnarray}
where we define [see also Eq.~(\ref{eq:many_covariant_derivatives})]
\begin{equation}
[\overleftrightarrow{\bm{D}}]^3
= \frac{1}{4} 
\left[ 
\overleftrightarrow{\bm{D}}
(\overleftrightarrow{\bm{D}})^2
+ 2
\overleftrightarrow{\bm{D}}^i
\overleftrightarrow{\bm{D}}
\overleftrightarrow{\bm{D}}^i
+
(\overleftrightarrow{\bm{D}})^2
\overleftrightarrow{\bm{D}}
\right].
\end{equation}
The color-octet matrix elements for the $J^{PC}=1^{+-}$ case are 
obtained from the color-octet matrix elements for the $J^{PC}=1^{--}$ case in
Eq.~(\ref{eq:QQgfactorization}) by making the replacements $g_s \bm{E} \to g_s
\bm{B}$ and $g_s \bm{B} \to g_s \bm{E}$. 

From the $J^{PC} = 1^{+-}$ contributions to the $Q \bar Q$ and $Q \bar Q g$
amplitudes in Eqs.~(\ref{eq:QQamplitude}) and (\ref{eq:QQgamplitude}), we obtain the
short-distance coefficients 
\begin{subequations}
\label{eq:WC_axialvector}%
\begin{eqnarray}
\label{eq:WC_axialvector_a}%
c_1 &=&  \frac{e e_Q y_Q}{m} 
(\bm{\epsilon}_\gamma^* \times \hat{\bm{p}}_\gamma) 
\cdot \bm{\epsilon}^* (\lambda), \\
c_{\bm{D}^3} &=& -\frac{4}{5} \frac{e e_Q y_Q}{m} 
(\bm{\epsilon}_\gamma^* \times \hat{\bm{p}}_\gamma) 
\cdot \bm{\epsilon}^* (\lambda), \\
c_{E} &=& i \frac{e e_Q y_Q}{m} 
(\bm{\epsilon}_\gamma^* \times \hat{\bm{p}}_\gamma) 
\cdot \bm{\epsilon}^* (\lambda), \\
c_{DB_0} &=& -i \frac{e e_Q y_Q}{m} 
(\bm{\epsilon}_\gamma^* \times \hat{\bm{p}}_\gamma) 
\cdot \bm{\epsilon}^* (\lambda)
\frac{1-3 r}{8(1-r)}, \\
c_{DB_1} &=& -i \frac{e e_Q y_Q}{m} 
(\bm{\epsilon}_\gamma^* \times \hat{\bm{p}}_\gamma) 
\cdot \bm{\epsilon}^* (\lambda)
\frac{3 - 7 r}{16 (1-r)}, \\
c_{DB_1'} &=& i \frac{e e_Q y_Q}{m} 
(\bm{\epsilon}_\gamma^* \times \hat{\bm{p}}_\gamma) 
\cdot \bm{\epsilon}^* (\lambda)
\frac{39 - 9 r}{40 (1-r)}, \\
c_{DB_2} &=& -i \frac{e e_Q y_Q}{m} 
(\bm{\epsilon}_\gamma^* \times \hat{\bm{p}}_\gamma) 
\cdot \bm{\epsilon}^* (\lambda)
\frac{31 + 9 r}{80 (1-r)}, 
\end{eqnarray}
\end{subequations}
where $r= \frac{4 m^2}{m_H^2}$. 

Recently, a computation of the decay rate $\Gamma(H \to h_c + \gamma)$ 
in the NRQCD factorization formalism at leading order in $v$ appeard in 
Ref.~\cite{Mao:2019hgg}. 
This is equivalent to our calculation of the short-distance coefficient $c_1$
in Eq.~(\ref{eq:WC_axialvector_a}), which leads to the expression for the decay
rate $\Gamma(H \to h_c + \gamma)$ at leading order in $v$ that is given by 
\begin{equation}
\label{eq:hcrate}
\Gamma(H \to h_c + \gamma) 
= \frac{4 \pi \alpha e_Q^2 y_Q^2}{3 m^4} 
\sum_{\lambda=0, \pm 1}
\langle 0 | 
 \chi^\dag 
(-\tfrac{i}{2} \overleftrightarrow{\bm{D}})^i 
\psi | h_c (\lambda) \rangle 
\langle h_c (\lambda) | \psi^\dag 
(-\tfrac{i}{2} \overleftrightarrow{\bm{D}})^i 
\chi | 0 \rangle 
\Phi,
\end{equation}
where the sum is over the polarization of the $h_c$ and $\Phi$ is the 
phase-space and normalization factor given in Eq.~(\ref{eq:phase_space}). 
Our result in Eq.~(\ref{eq:hcrate}) agrees with the decay rate
computed in Ref.~\cite{Mao:2019hgg}.

\section{Gremm-Kapustin relations}

The Gremm-Kapustin relations~\cite{Gremm:1997dq} are obtained from
\begin{equation}
\langle V | [{\cal O}, H ] | 0 \rangle
= 
- \langle V | H {\cal O} | 0 \rangle
= 
(2 m - m_V)  \langle V | {\cal O} | 0 \rangle, 
\end{equation}
where ${\cal O}$ is an NRQCD operator, and $H$ is the NRQCD Hamiltonian. 
Computing the commutator $[{\cal O}, H]$ leads to the following relations:
\begin{subequations}
\label{eq:Gremm}%
\begin{eqnarray}
&& \hspace{-10ex} 
(m_V - 2 m) 
\langle V | \psi^\dag \bm{\sigma} \cdot \bm{\epsilon}(\lambda) \chi | 0 \rangle 
\nonumber \\
&=& \frac{1}{m} 
 \langle V | \psi^\dag \bm{\sigma} \cdot \bm{\epsilon} (\lambda)
(-\tfrac{i}{2} \overleftrightarrow{\bm{D}})^2 \chi | 0 \rangle
\nonumber \\ && 
- \frac{1}{4 m^3} \langle V | \psi^\dag \bm{\sigma}
\cdot \bm{\epsilon} (\lambda)
(-\tfrac{i}{2} \overleftrightarrow{\bm{D}})^4 \chi | 0 \rangle
\nonumber \\ && 
- \frac{1}{m} \langle V | \psi^\dag g_s \bm{B} \cdot \bm{\epsilon} (\lambda)
\chi | 0 \rangle 
\nonumber \\ && 
+ \frac{1}{4 m^2} 
\langle V | \psi^\dag \bm{\epsilon} (\lambda)  \cdot 
\tfrac{1}{2} 
[\bm{\sigma} \times ( \overleftrightarrow{\bm{D}} \times g_s \bm{E}
- g_s \bm{E} \times \overleftrightarrow{\bm{D}} )] \chi | 0 \rangle , 
\\
&& \hspace{-10ex}
(m_V - 2 m) 
 \langle V | \psi^\dag \bm{\sigma} \cdot \bm{\epsilon} (\lambda)
(-\tfrac{i}{2} \overleftrightarrow{\bm{D}})^2 \chi | 0 \rangle
\nonumber \\
&=& 
\frac{1}{m} 
\langle V | \psi^\dag \bm{\sigma} \cdot \bm{\epsilon} (\lambda)
(-\tfrac{i}{2} \overleftrightarrow{\bm{D}})^4 \chi | 0 \rangle
\nonumber \\ &&
- \frac{3}{2} 
\langle V | \psi^\dag 
\bm{\sigma} \cdot \bm{\epsilon} (\lambda) 
\tfrac{1}{3} 
( \overleftrightarrow{\bm{D}} \cdot g_s
\bm{E} 
+g_s \bm{E} \cdot \overleftrightarrow{\bm{D}} ) 
\chi | 0 \rangle, 
\\ 
&& \hspace{-10ex}
\label{eq:Gremm3}%
(m_V - 2 m) 
\langle V | \psi^\dag 
\epsilon^i (\lambda) \sigma^j (-\tfrac{i}{2})^2 
\overleftrightarrow{\bm{D}}^{(i} \overleftrightarrow{\bm{D}}^{j)} 
\chi | 0 \rangle 
\nonumber \\
&=& 
\frac{1}{m} 
\langle V | \tfrac{1}{2} 
\psi^\dag \epsilon^i (\lambda) \sigma^j (-\tfrac{i}{2})^2 
\{\overleftrightarrow{\bm{D}}^{(i} \overleftrightarrow{\bm{D}}^{j)}, 
(-\tfrac{i}{2} \overleftrightarrow{\bm{D}} )^2 \}
\chi | 0 \rangle 
\nonumber \\ && 
- \frac{1}{2} 
\langle V | \psi^\dag \epsilon^i (\lambda) \sigma^j 
( \overleftrightarrow{\bm{D}}^{(i} g_s \bm{E}^{j)} 
+ g_s \bm{E}^{(i} \overleftrightarrow{\bm{D}}^{j)} )
\chi | 0 \rangle, 
\\
&& \hspace{-10ex}
\label{eq:Gremm4}%
(m_V - 2 m) 
\langle V | \psi^\dag g_s \bm{B} \cdot \bm{\epsilon} (\lambda) \chi | 0 \rangle 
\nonumber \\
&=& \langle V | \psi^\dag 
\bm{\epsilon} (\lambda) \cdot 
\tfrac{i}{2} 
( \overleftrightarrow{\bm{D}} \times g_s \bm{E} 
+g_s \bm{E} \times \overleftrightarrow{\bm{D}} ) \chi | 0 \rangle , 
\end{eqnarray}
\end{subequations}
where, in calculating the commutator $[{\cal O},H]$ 
we included operators in the Hamiltonian up to $1/m^2$ accuracy, 
with the Wilson coefficients at order $\alpha_s^0$, 
and kept NRQCD operators up to dimension 7.
Hence, 
the relations in Eqs.~(\ref{eq:Gremm}) are valid up to order $v^4$ relative to 
the leading-order LDME $\langle V | \psi^\dag \bm{\sigma} \cdot 
\bm{\epsilon}(\lambda) \chi | 0 \rangle$ and at leading order in $\alpha_s$. 
These relations can also be verified in perturbation theory.

The Gremm-Kapustin relations provide a way to identify the
velocity scalings of the LDMEs that are suppressed beyond the conservative
power counting of Refs.~\cite{Brambilla:2001xy, Brambilla:2002nu,
Brambilla:2006ph, Brambilla:2008zg}.
Since the binding energy $m_V - 2 m$ scales like $m v^2$, 
the left-hand side of Eq.~(\ref{eq:Gremm3}) is suppressed by 
$v^6$ compared to the leading-order LDME 
$\langle V | \psi^\dag \bm{\sigma} \cdot \bm{\epsilon}(\lambda) \chi | 0
\rangle$. Therefore, the LDME $\langle V | \psi^\dag \epsilon^i (\lambda)
\sigma^j ( \overleftrightarrow{\bm{D}}^{(i} g_s \bm{E}^{j)} 
+ g_s \bm{E}^{(i} \overleftrightarrow{\bm{D}}^{j)} )
\chi | 0 \rangle$ does not contribute to the amplitude in
Eq.~(\ref{eq:Jpsifactorization}) at relative order $v^4$ accuracy 
because it is suppressed by at least $v^6$
compared to the leading-order LDME, and scales like $v^{15/2}$. 
Similarly, the left-hand side of Eq.~(\ref{eq:Gremm4}) is suppressed by $v^5$
compared to the leading-order LDME, and hence, the LDME on the right-hand side
of Eq.~(\ref{eq:Gremm4}) scales like $v^{13/2}$ and 
does not contribute to the amplitude in
Eq.~(\ref{eq:Jpsifactorization}) at relative order $v^4$ accuracy.

 
\end{document}